\documentclass[11pt]{article}
\usepackage{graphicx}
\usepackage{amsfonts}
\usepackage{amssymb}
\usepackage{amsmath}
\usepackage{bbm}
\usepackage[usenames, pdftex, dvipsnames]{color}
\usepackage{url}
\usepackage{umoline}

\newtheorem{theorem}{Theorem}[section]

\newtheorem{deff}[theorem]{Definition}  
\newtheorem{definition}[theorem]{Definition}

\newtheorem{conj}[theorem]{Conjecture}

\newtheorem{corol}[theorem]{Corollary}

\newcommand{\Fig}[1]{Fig.~\ref{#1}}

\newcommand{\Sec}[1]{Sec.~\ref{#1}}
\newcommand{\Ref}[1]{Ref.~\cite{#1}}

\newcommand{\QMA}{\mathsf{QMA}}
\newcommand{\MA}{\mathsf{MA}}

\newcommand{\NP}{\mathsf{NP}}
\newcommand{\PSPACE}{\mathsf{PSPACE}}

\newcommand{\NEXP}{\mathsf{NEXP}}

\newcommand{\IP}{\mathsf{IP}}
\newcommand{\QIP}{\mathsf{QIP}}
\newcommand{\MIP}{\mathsf{MIP}}
\newcommand{\QMIP}{\mathsf{QMIP}}
\newcommand{\NLTS}{\mathrm{NLTS}}
\newcommand{\CLH}{\mathsf{CLH}}
\newcommand{\qLTC}{\mathsf{qLTC}}
\newcommand{\LTC}{\mathsf{LTC}}
\newcommand{\PCP}{\mathrm{PCP}}
\newcommand{\CSP}{\mathrm{CSP}}
\newcommand{\qPCP}{\mathrm{qPCP}}

\newcommand{\Id}{{\mathbb{I}}}

\newcommand{\ket}[1]{{ |{#1} \rangle }}  
\newcommand{\bra}[1]{{ \langle {#1} | }}
\newcommand{\braket}[2]{{ \langle {#1} | {#2} \rangle}}

\newcommand{\Oorderof}{\mathcal{O}}
\newcommand{\orderof}[1]{\Oorderof(#1)} 
\newcommand{\poly}{\mathrm{poly}} 
\newcommand{\EqDef}{:=}

\newcommand{\Tr}{\mathop{\rm Tr}\nolimits}

\newcommand{\UNSAT}{\mathrm{UNSAT}}
\newcommand{\QUNSAT}{\mathrm{QUNSAT}}
\newcommand{\SAT}{\mathrm{SAT}}

\newcommand{\Av}{\mathbb{E}}

\newcommand{\ignore}[1]{{}}

\newcommand{\eps}{\varepsilon}  
 \newcommand{\ra}{\rangle}

 \newcommand{\qedsymb}{\hfill{\rule{2mm}{2mm}}}

\newcommand{\MC}{\mathcal{C}}

\newcommand{\gs}{\Omega}

\newcommand{\Egs}{E_0}

\newcommand{\Gap}{\Gamma}
\newcommand{\gap}{\gamma}

\newcommand{\bran}{r}

\begin{document}

\title{The Quantum PCP
Conjecture}
\author{Dorit Aharonov\footnote{School of Engineering and Computer Science,
  The Hebrew University, Jerusalem, Israel.} \ 
and Itai Arad\footnote{Centre for Quantum Technologies, 
  National University of Singapore, Singapore.}\  \  
and Thomas Vidick\footnote{Computer Science and Artificial 
  Intelligence Laboratory, Massachusetts Institute of 
  Technology, USA and Centre for Quantum Technologies, 
  National University of Singapore, Singapore.}}
\date{\today
}
\maketitle
\begin{abstract}
  The classical $\PCP$ theorem is arguably the most important
  achievement of classical complexity theory in the past quarter
  century. In recent years, researchers in quantum computational
  complexity have tried to identify approaches and develop tools
  that address the question: does a quantum version of the $\PCP$
  theorem hold?  The story of this study starts with classical
  complexity and takes unexpected turns providing fascinating vistas
  on the foundations of quantum mechanics, the global nature of
  entanglement and its topological properties, quantum error
  correction, information theory, and much more; it raises questions
  that touch upon some of the most fundamental issues at the heart
  of our understanding of quantum mechanics.  At this point, the
  jury is still out as to whether or not such a theorem holds. This
  survey aims to provide a snapshot of the status in this ongoing
  story, tailored to a general theory-of-CS audience.  
\end{abstract}

\section{Introduction}
\label{sec:introduction}

Perhaps the most fundamental result in classical complexity theory
is the Cook-Levin theorem \cite{ref:Cook71, ref:Levin73}, which
states that $\SAT$, the problem of deciding satisfiability of a
Boolean formula, is $\NP$-complete. This result opened the door to
the study of the rich theory of $\NP$-completeness of constraint
satisfaction problems ($\CSP$s). At the heart of this framework
stands the basic understanding that computation is local, made of
elementary steps which can be verified one at a time. 

The main object of this study is the $k$-local constraint
satisfaction problem.  A $k$-$\CSP$ is a formula on $n$ Boolean (or
over a larger alphabet) variables, composed of $m$ constraints, or
clauses, each acting on at most $k$ variables, where $k$ should be
thought of as a small constant (say, $2$ or $3$). By a constraint,
we mean some restriction on assignments to the $k$ variables which
excludes one or more of the $2^k$ possibilities.  As a consequence
of the Cook-Levin theorem, deciding whether or not a $\CSP$ instance
has a satisfying assignment is exactly as hard as deciding whether a
given polynomial-time Turing machine has an accepting input: it is
$\NP$-complete.

Starting with the development of interactive proofs in the 1980s, a
long line of work in complexity theory has resulted in a
considerable strengthening of the Cook-Levin theorem, leading to 
the celebrated $\PCP$ (for Probabilistically Checkable Proofs) 
theorem~\cite{ref:PCP1, ref:PCP2}. In its gap amplification version
due to Dinur~\cite{ref:Dinur-PCP}, the $\PCP$ theorem states that it
is $\NP$-hard to distinguish between the cases when an instance of
$2$-$\CSP$ is completely satisfiable, or when no more than $99\%$ of
its constraints can be satisfied.  In other words, not only is it
$\NP$-hard to determine exactly whether all clauses are
simultaneously satisfiable or any assignment violates at least one
clause, but it remains $\NP$-hard to do so when one is promised that
any assignment must either satisfy all clauses or violate a constant
fraction of clauses. In fact, a major development stemming from the
$\PCP$ theorem is research on \emph{hardness of approximation},
where one attempts to determine for which approximation factors a
given class of $k$-$\CSP$s remains $\NP$-hard. A surprising outcome
of this line of work has been that for many $k$-$\CSP$s, the
hardness of approximation factor matches that achieved by a random
assignment. For instance, a random assignment to a $3$-$\SAT$
formula already satisfies $7/8$ of the clauses in expectation, and
it is $\NP$-hard to do even slightly better, namely to distinguish
formulas for which at most a fraction $7/8+\eps$ of clauses can be
simultaneously satisfied from formulas which are fully satisfiable.

The original version of the $\PCP$ theorem \cite{ref:PCP1, ref:PCP2}
was stated quite differently.  Owing to its origins in the
development of the celebrated 
$\IP=\PSPACE$~\cite{LunForKarNis92JACM,ref:Sha92} and
$\MIP=\NEXP$~\cite{BabForLun91CC} results from the theory of
interactive proofs \cite{ref:Babai-IP,ref:Goldwasser-IP}, it was
initially formulated as follows: any language in $\NP$ can be
verified, up to a constant probability of error, by a randomized
polynomial-time verifier who only reads a constant (!)~number of
(randomly chosen) bits from a polynomial-size proof. Hence the term
\emph{probabilistically checkable proofs}. Though this formulation
may a priori sound quite different from the gap amplification one
described above, it is quite easy to see they are
equivalent~\cite{ref:AroraBarak}: roughly, if any assignment must
violate a constant fraction of the clauses, then sampling a clause
at random and checking whether it is satisfied (which only requires
reading the bits corresponding to the $k$ variables on which it
acts) would detect a violation with constant probability.  It is
often fruitful to go back and forth between these two pictures; we
will make use of both here. 

\subsection{Quantum Hamiltonian complexity}
\label{sec:QHC}

Over the past decade, a fascinating analogy has been drawn between 
the above major thread in classical computational complexity, namely
the study of $\CSP$s, and the seemingly unrelated field of condensed
matter physics. The object of condensed matter physics is the study
of properties of condensed phases of matter, such as solids or
liquids, in which systems typically consist of many interacting
particles, governed by the laws of quantum or classical mechanics. A
central question of interest is which configurations of the
particles minimize the energy, and what this minimal energy is.  The
energy of the system is determined by an operator called the
\emph{Hamiltonian}. It typically consists of a sum of \emph{local}
terms, namely terms that determine the energy of a small number of
``neighboring'' particles.\footnote{While in practice these terms
are often localized in space, here, unless explicitly stated
otherwise, by ``local'' we shall mean involving $\orderof{1}$
particles which can be arbitrarily far away from each other.} The
total energy is the sum of contributions coming from each of those
local terms. We can think of each local term as a generalized
constraint, and its contribution to the total energy as a signature
of how violated the constraint is. The question of finding the
configuration of lowest energy has a very similar flavor to the
central question in $\CSP$s: What is the assignment that violates
the fewest clauses?  We first introduce the mathematical formalism
associated with the description of local Hamiltonians, and then
explain how this connection can be made precise by showing how
$k$-$\CSP$s can be viewed as special instances of local
Hamiltonians.

We first need to describe the state space in quantum many-body
physics.\footnote{Throughout this column we aim to provide the
reader with the necessary background on quantum computation.  For a
more in-depth introduction, see for example
\cite{ref:NielsenChuang}.} The space of \emph{pure} states of $n$
two-state particles, also called quantum bits, or \emph{qubits}, is
a complex vector space of dimension $2^n$. It is spanned by an
orthonormal basis of $2^n$ pure states, which we call the
\emph{computational basis} and denote by
$|i_1,...,i_n\ra=|i_1\ra\otimes\cdots\otimes|i_n\ra$, where $i_j\in
\{0,1\}$. The notation $\ket{\cdot}$, called \emph{Dirac notation},
provides a way to clarify that we are speaking of a column vector,
also called \emph{ket}, in a Hilbert space.  The \emph{bra} notation
$\bra{\phi}$ is used to denote a row vector; $\braket{\phi}{\psi}$
denotes the inner (scalar) product, while $\ket{\psi}\bra{\phi}$
denotes the outer product, a rank-1 matrix. The reader unfamiliar
with the definition of the tensor product $\otimes$ may treat it as
a simple notation here; it possesses all the usual properties of a
product --- associativity, distributivity and so on.  States, such
as the basis states, which can be written as a tensor product of
single-qubit states are called \emph{product states}.  States which
cannot be written as tensors of single-qubit states are called
\emph{entangled}, and they will play a crucial role later
on.\footnote{Note that it is not always trivial to determine whether
a state is entangled or not. For instance, the state 
$1/2(\ket{00}-\ket{01}+\ket{10}-\ket{11})$ is \emph{not} entangled:
it is the product state $((\ket{0}+\ket{1})/\sqrt{2})\otimes
((\ket{0}-\ket{1})/\sqrt{2})$.} A general pure state of the $n$
qubits is specified by a vector
$|\psi\ra=\sum_{i}a_i|i_1,...,i_n\ra$, where $a_i$ are complex
coefficients satisfying the normalization condition $\sum_i
|a_i|^2=1$. Such a linear combination of basis states is also called
a \emph{superposition}. All the above can be generalized to higher
$d$-dimensional particles, often called \emph{qudits}, in a
straightforward manner. 

A $k$-local Hamiltonian $H$ acting on a system of $n$ qudits is a
$d^n \times d^n$ matrix that can be written as a sum $H=\sum_{i=1}^m
H_i$, where each $H_i$ is Hermitian of operator norm $\|H_i\|\leq 1$
and acts non-trivially only on $k$ out of the $n$ particles.
Formally, this means that $H_i$ can be written as the tensor product
of some matrix acting on $k$ qubits and the identity on the
remaining qubits. For the purposes of this column it is simplest to
think of each $H_i$ as a projection; the general case does not
introduce any new essential properties.  Given a state $\ket{\psi}$,
its \emph{energy} with respect to the Hamiltonian $H$ is defined to
be $\bra{\psi}H\ket{\psi} = \sum_{i=1}^m\bra{\psi}H_i\ket{\psi}$.
How to interpret the value $\bra{\psi}H_i\ket{\psi}$? When $H_i$ is
a projection then $H_i^2=H_i$; in this case it is exactly the norm
squared of the vector $H_i\ket{\psi}$. By the laws of quantum
mechanics this is exactly the \emph{probability of obtaining the
outcome} `$H_i$' when measuring the state with respect to the
two-outcome measurement $\{H_i,\Id -H_i\}$, and we can think of it
as the probability that $\ket{\Psi}$ \emph{violates} the $i$-th
constraint. We note that if $\ket{\psi}$ is an eigenstate (an
eigenvector) of $H$, then $\bra{\psi}H\ket{\psi}$ is the eigenvalue
associated with $\ket{\psi}$.  The eigenvalues of $H$ are called the
\emph{energy levels} of the system. The lowest possible energy level
(the \emph{ground energy}) is the smallest eigenvalue of $H$ and is
denoted $\Egs$; the corresponding eigenstate (or eigenspace) is
called the \emph{groundstate} (or groundspace). Computing the ground
energy and understanding the characteristics of the groundstates for
a variety of physical systems is arguably the subject of the
majority of the theoretical research in condensed matter physics. 

Kitaev introduced a computational viewpoint on those physical
questions by defining the following
problem~\cite{ref:Kit99,ref:KitaevShenVyalyi}.

\begin{deff}[The \boldmath{$k$}-local Hamiltonian (LH) problem] \ 
\label{def:LH}
  \begin{itemize}
    \item  \textbf{Input:} $H_1,\ldots,H_m$, a set of $m$ Hermitian matrices
      each acting on $k$ qudits  
      out of an $n$-qudit system and
      satisfying $\|H_i\|\le 1$. Each matrix entry is specified by
      $\poly(n)$-many bits. Apart from the $H_i$ we are also given
      two real numbers, $a$ and $b$ (again, with polynomially many
      bits of precision) such that $\Gap\EqDef b-a>1/poly(n)$.
      $\Gap$ is referred to as the \emph{absolute promise gap} of
      the problem; $\gap\EqDef\Gap/m$ is referred to as the
      \emph{relative promise gap}, or simply the promise gap.

    \item \textbf{Output:} Is the smallest eigenvalue 
      of $H=H_1+H_2+...+H_m$ smaller than $a$ or are all its
      eigenvalues larger than $b$?\footnote{Note that this is
      equivalent to asking whether there exists a state $\ket{\psi}$
      such that the expectation value $\bra{\psi}H\ket{\psi}\leq a$,
      or $\bra{\psi}H\ket{\psi}\geq b$ for all states $\ket{\psi}$.}
  \end{itemize} 
\end{deff}

We indicate why the LH problem is a natural generalization of
$\CSP$s to the quantum world by showing how $3$-$\SAT$ can be
viewed as a $3$-local Hamiltonian problem.  Let $\phi = C_1 \wedge
C_2 \wedge \dots \wedge C_m$ be a $3$-$\SAT$ formula on $n$
variables, where each $C_i$ is a disjunction over three variables or
their negations. For every clause $C_i$, introduce an $8 \times 8$
matrix $H_i$ acting on three qubits, defined as the projection on
the computational basis state associated with the unsatisfying
assignment of $C_i$. For example, for the clause $C_i=X_1 \vee X_2
\vee \neg X_3$ we obtain the matrix
\begin{align*}
H_i=     \left ( \begin{array}{cccccccc}        
                0&0&0&0&0&0&0&0\\
                0&1&0&0&0&0&0&0\\
                0&0&0&0&0&0&0&0\\
                0&0&0&0&0&0&0&0\\
                0&0&0&0&0&0&0&0\\
                0&0&0&0&0&0&0&0\\
                0&0&0&0&0&0&0&0\\
                0&0&0&0&0&0&0&0\\
               \end{array}
\right ).
\end{align*}
We then extend $H_i$ to an operator on all the qubits by taking its
tensor product with the identity on the remaining qubits. In a
slight abuse of notation we denote the new matrix by $H_i$ as well.
If $z$ is an assignment to the $n$ variables, which satisfies a
clause $C_i$, then $H_i|z\ra = 0$, namely, $\ket{z}$ has $0$ energy
with respect to that clause.  Otherwise, $H_i |z\ra = |z\ra$, and so
the energy of $\ket{z}$ with respect to that clause is $1$. Denoting
$H = \sum_{i=1}^{m} H_i$, it is now clear that $H|z\ra = q|z\ra$
where $q$ is the number of clauses violated by $z$. This gives an
eigenbasis of $2^n$ orthonormal eigenstates for $H$, corresponding
to the $2^n$ classical strings of $n$ bits, and any assignment $z_0$
which violates the least number of clauses gives rise to a
groundstate $\ket{z_0}$ of minimal energy with respect to $H$.  We
note that in this case, since we found a basis of eigenstates which
are all tensor product states, entanglement does not a play a role
in this problem. Our construction thus shows that $3$-$\SAT$ is
equivalent to the problem: ``Is the smallest eigenvalue of $H$ at
most $0$, or is it at least $1$?'', and therefore it is an instance
of the $3$-local Hamiltonian problem. 

We have shown that $\CSP$s can be seen as a very special class of
local Hamiltonians: those for which all the local terms are diagonal
in the computational basis. In this case, the eigenstates of the
system are the computational basis states, which are simple product
states with no entanglement.  However, the local Hamiltonians that
are considered in condensed matter physics are far more general: the
local terms need not be diagonal in the computational basis.  The
groundstates of such Hamiltonians may contain intricate forms of
entanglement, spread across all particles.  The presence of this
entanglement is at the core of the challenges that the field of
quantum Hamiltonian complexity, which attempts to provide a
\emph{computational perspective} on the study of local Hamiltonians
and their groundstates, faces. 

\subsection{The quantum PCP conjecture}
\label{sec:qpcpconj}

A central effort of quantum Hamiltonian complexity is to develop
insights on multipartite entanglement by understanding how its
presence affects classical results on the complexity of $\CSP$s. 
The first step in this endeavor has already been completed to a
large extent.  In 1999 Kitaev \cite{ref:Kit99,ref:KitaevShenVyalyi}
established the quantum analogue of the Cook-Levin theorem. First,
he formally defined the quantum analogue of $\NP$, called $\QMA$
(apparently, this notion was first discussed by
Knill~\cite{ref:Knill96}). $\QMA$ is defined just like $\MA$ (the
probabilistic analogue of $\NP$) except both the witness and
verifier are quantum: 
\begin{definition}[The complexity class $\QMA$] 
\label{def:QMA} 
  A language $L\subseteq \{0,1\}^*$ is in $\QMA$ if there exists a
  quantum polynomial time algorithm $V$ (called the \emph{verifier})
  and a polynomial $p(\cdot)$ such that:
  \begin{itemize}
    \item $\forall x \in L$ there exists a state $|\xi\ra$  
      on $p(|x|)$ qubits such that $V$ accepts the pair of inputs
      $(x,\ket{\xi})$ with probability at least $2/3$.
      
    \item $\forall x \notin L$ and for all states $|\xi\ra$ on 
      $p(|x|)$ qubits, $V$ accepts $(x,\ket{\xi})$ with probability
      at most $1/3$. 
  \end{itemize}
\end{definition}
Kitaev showed that the LH problem is complete for $\QMA$. The fact
that it is inside $\QMA$ is quite simple (the witness is expected to
be the lowest energy eigenstate); the other direction is more
involved and proving it requires overcoming obstacles unique to the
quantum setting.  The main ingredient is the circuit-to-Hamiltonian
construction, which maps a quantum circuit to a local Hamiltonian
whose groundstate encodes the correct history of the computation of
the circuit; we return to this construction in
Section~\ref{sec:cooklevin}.  Many exciting results build on 
Kitaev's construction. To mention a few: the equivalence of quantum
computation and adiabatic quantum computation~\cite{ref:adiabatic},
extensions of $\QMA$-hardness to restricted classes of
Hamiltonians~\cite{ref:adiabatic, ref:KempeKitaevRegev,
ref:TerhalOliveira}, including the counter-intuitive discovery that
even local Hamiltonians acting on nearest-neighbor particles
arranged on a line are $\QMA$-hard~\cite{ref:QMA-1D}, even for
translationally invariant systems~\cite{ref:GottesmanIrani}; the
invention of quantum gadgets~\cite{ref:KempeKitaevRegev,
ref:TerhalOliveira, ref:Terhal-Gadgets}, which resemble classical
graphical gadgets often used in classical $\NP$
reductions~\cite{ref:Karp}; and a proof that a famous problem in
physics (known as the ``universal functional'' problem, and related
to finding the groundstate of electrons interacting through the
Coulomb potential) is $\QMA$-hard~\cite{ref:densityFunc}.

However, viewed from a wider perspective, the current situation in
quantum Hamiltonian complexity can be compared to the situation in
classical computational complexity in the early $1970$s: after the
foundations of the theory of $\NP$-hardness had been laid down by
Cook, Levin \cite{ref:Cook71,ref:Levin73} and Karp \cite{ref:Karp},
but before the fundamental breakthroughs brought in by the study of
interactive proofs and the $\PCP$ theorem.  Does a quantum version
of the $\PCP$ theorem hold?  The question was first raised more than
a decade ago \cite{ref:AharonovNaveh}, then several times later
(e.g., in \cite{ref:AaronsonQPCP}), and rigorously formulated in
\cite{ref:qgap-amp}.  Loosely speaking, while the quantum Cook-Levin
theorem asserts that it is $\QMA$-hard to approximate the ground
energy of a local Hamiltonian up to an additive error of $\Gamma
=1/\poly$, the $\qPCP$ conjecture states that it remains $\QMA$-hard
to approximate it even up to an error $\Gamma = \gamma m$, where
$0<\gamma<1$ is some constant and $m$ is the number of local terms
in the Hamiltonian.

\begin{conj} [Quantum PCP by gap amplification] 
\label{con:qgap} 
  The LH problem with a constant relative promise gap $\gap>0$ is $\QMA$-hard
  under quantum polynomial time reductions. 
\end{conj}  

More formally, the conjecture states that there exists a quantum
polynomial time transformation that takes any local Hamiltonian
$H=\sum_{i=1}^m H_i$ with absolute promise gap
$\Gap=\Omega(1/\poly)$, and, with constant non-zero probability, 
transforms it into a new local Hamiltonian $H'=\sum_{i=1}^{m'} H'_i$
with an absolute promise gap $\Gap'=\Omega(m')$, such that if the
original Hamiltonian had ground energy at most $a$ (at least $b\geq
a+\Gap$) then the new Hamiltonian has ground energy of at most $a'$
(at least $b'\geq a'+\Gap'$).  Just as with the classical $\PCP$
theorem, the above conjecture has an equivalent statement in terms
of efficient proof verification. 

\begin{conj}[Quantum PCP by proof verification]
\label{con:randomaccessqpcp} 
  For any language in $\QMA$ there exists a polynomial time quantum
  verifier, which acts on the classical input string $x$ and a
  witness $\ket{\xi}$, a quantum state of $poly(|x|)$ qubits, such
  that the verifier accesses only $\orderof{1}$ qubits from the
  witness and decides on acceptance or rejection with constant error
  probability.  
\end{conj}

The proof that the two $\qPCP$ conjectures are equivalent is not 
difficult \cite{ref:qgap-amp}, but to the best of our knowledge the
fact that the reductions in the statement of
Conjecture~\ref{con:qgap} are allowed to be quantum is necessary for
the equivalence to hold. More precisely, this is needed to show that
Conjecture~\ref{con:randomaccessqpcp} implies 
Conjecture~\ref{con:qgap}.\footnote{\label{footnote:direction} The
verifier of Conjecture~\ref{con:randomaccessqpcp} is translated to a
local Hamiltonian in the following way.  For each $k$-tuple of
qubits that the quantum verifier reads from the random access
quantum proof, there is a $k$-local term in the Hamiltonian which
checks that the computation of the verifier on those $k$ qubits
accepts.  To compute the local term, however, one needs to simulate
efficiently the action of the verifier, a quantum polynomial time
algorithm.} The fact that the simple connection between the two
equivalent formulations of the classical $\PCP$ theorem already becomes
much more subtle in the quantum world may hint at the possible
difficulties in proving a $\qPCP$ conjecture. In most of this column
we shall refer to Conjecture~\ref{con:qgap} as the $\qPCP$
conjecture.

\subsection{Physical underpinnings for the conjecture}

In addition to it being a natural and elegant extension of the
$\PCP$ theorem, interest in the $\qPCP$ conjecture is many-fold.
First, as we will explain in great detail, the heart of the
difficulty of the $\qPCP$ conjecture is the presence of exotic forms
of multipartite entanglement in the groundstates of local
Hamiltonians. It is this entanglement which makes the conjecture a
much more challenging extension of the classical $\PCP$ theorem than
it might seem at first sight. It also provides great motivation for
the study of the conjecture, an excellent computational probe which
may lead to new insights into some of the most counter-intuitive
phenomenon of quantum mechanics. A significant part of this paper
is devoted to explaining this aspect of the $\qPCP$ conjecture.  

A Hamiltonian's groundstate, however, is not the only state of
interest. In fact, one may argue that this state is not ``natural''
in that it only describes the state of the system in the limit of
infinitely low temperatures.  An additional motivation for the study
of $\qPCP$ is its direct relevance to the stability of entanglement
at so-called ``room temperature''.  Indeed, while physical intuition
suggests that quantum correlations typically do not persist at
finite non-zero temperatures for systems in equilibrium,\footnote{
Though there exists an example -- the 4D toric code -- which already
demonstrates that this intuition can be violated at least to some
extent; see \Sec{sec:NLTS} and \Ref{ref:nonzeroTQO}.} the $\qPCP$
conjecture implies exactly the opposite! The gist of the argument is
that the $\QMA$-hardness of providing even very rough approximations
to the groundstate energy, as asserted by the $\qPCP$ conjecture, 
implies that no low-energy state can have an efficient classical
description (and hence must be entangled): such a description would
automatically lead to a classical witness for the energy.
 
More precisely, the argument goes as follows.  A physical system at
equilibrium at temperature $T>0$ is described by the so-called
Gibbs-Boltzmann distribution over eigenstates $\ket{E_i}$, with
corresponding energy $E_i$, of its Hamiltonian.  Up to
normalization, this distribution assigns probability $e^{-E_i/T}$ to
the eigenstate $\ket{E_i}$; in the limit $T\to 0$ the system is
concentrated on its groundstate, which motivates the special role
taken by groundstates in the study of physical systems. 

But what happens at higher temperatures? 
Since the probabilities associated to the energies decay
exponentially fast, the total contribution of states with energy
above $\Egs+\Theta$, for some $\Theta \sim n T$, is exponentially
small. Suppose the $\qPCP$ holds for some absolute promise gap
$\Gap=\Omega(m)=\Omega(n)$, with $m$ being the number of
constraints, which we may assume to be larger than $n$. Then all
states of energy below $\Egs+\Gap$ must be highly entangled, for
otherwise one of them could be provided as a classical witness for
the ground energy of $H$, putting the problem in
$\NP$.\footnote{Note that for this to be true we are implicitly 
taking as part of the definition of a state with ``low
entanglement'' that such a state can be given a polynomial-size
classical representation from which one can efficiently compute the
energy. Only for states in 1D do we know that low entanglement, as
measured e.g. by the von Neumann entropy, implies such a
representation (using so-called \emph{matrix product
states}~\cite{PerezMPS07}).} By taking $T$ to be a small enough
constant we can make $\Theta<\Gap$, hence for such Hamiltonians the
Gibbs-Boltzmann distribution at constant temperature $T>0$ is
supported on highly entangled states (up to an exponentially small 
correction), contradicting the physical intuition. This suggests
that resolving the $\qPCP$ conjecture might shed light on the kind
of Hamiltonians which possess robust entanglement at room
temperature, and what physical properties such Hamiltonians must
exhibit. 

We mention that one can draw an even stronger statement from the
$\qPCP$ conjecture: that quantum entanglement must play a crucial
role in the dynamics of reaching equilibrium even at room
temperature, for certain quantum systems, assuming $\QMA\ne\NP$.
This is because for the system to \emph{reach} its equilibrium at
room temperature, even approximately, a $\QMA$-hard problem must be
solved.\footnote{Indeed, a state drawn from the Gibbs distribution
at room temperature can serve as a witness to solve a $\QMA$-hard
problem, namely estimating the energy of the system's Hamiltonian. 
This generalizes the same Gibbs distribution argument applied to
classical systems: it follows from the classical $\PCP$ theorem that
for general classical systems to reach their equilibrium at room
temperature they must solve an $\NP$-hard problem.}

This apparent contradiction between physical intuition and the
$\qPCP$ conjecture is captured by an elegant conjecture due to
Hastings~\cite{ref:NLTS2}, the $\NLTS$ (no low-energy trivial
states) conjecture. The conjecture isolates the notion of robustness
of entanglement from the remaining difficulties related to the
$\qPCP$ conjecture: it states that Hamiltonians whose low-energy
states are all highly entangled do exist. While the $\NLTS$
conjecture must be true for the $\qPCP$ to hold, the other direction
is not known.  Much of the recent progress on the $\qPCP$ conjecture
can be phrased in terms of this conjecture, and we devote a whole
section of this survey to recent progress on the $\NLTS$ conjecture,
both negative and positive.

\subsection{Outlook}

Despite the recent flurry of results attempting to make progress on
the $\qPCP$ conjecture and related topics (e.g., \cite{ref:qgap-amp,
ref:nogo, ref:DoritLior1, ref:schuch, ref:NLTS1, ref:NLTS2,
ref:GharibianKempe, ref:Vid13, ref:BH, ref:DoritLior2,
ref:Hastings2D}), there does not seem to be a clear consensus
forming as to whether it is true or false.  But what is becoming
undoubtedly clear is that much like it was the case in the long
journey towards the proof of the classical $\PCP$
theorem~\cite{ref:PCP-history}, the study and attempts to resolve
the $\qPCP$ conjecture bring up beautiful new questions and points
of view, and the goal of this survey is to present some of these
developments. 

We proceed with a discussion of multipartite entanglement, including
the EPR state, CAT states and a fundamental example of the global
properties of multipartite entanglement, Kitaev's toric
code~\cite{ref:toric}.  In Section~\ref{sec:quantizing} we explain
how issues raised by multipartite entanglement are resolved in the
proof of the quantum Cook-Levin theorem, and what further obstacles
are posed when trying to extend classical proofs of the $\PCP$
theorem to the quantum domain. We also present a recent result of
Brand\~ao and Harrow \cite{ref:BH} which captures formally some
intrinsic limitations on attempts at proving the $\qPCP$ conjecture
by following the classical route.  In Section~\ref{sec:NLTS} we
introduce Hastings' NLTS conjecture regarding robust entanglement at
room temperature, explain in more detail its connection to the
$\qPCP$ conjecture, and describe recent results \cite{ref:BV,
ref:DoritLior1, ref:DoritLior2,ref:NLTS1} which provide strong
limitations on the Hamiltonians which could possibly possess such
robust entanglement; we also describe a recent positive attempt
\cite{ref:NLTS2} based on low-dimensional topology. In our last
section we take a look at the original line of works which led to
the proof of the $\PCP$ theorem, namely interactive proofs.  We
present an exponential size classical $\PCP$ for $\QMA$ (resolving
an open question from \cite{ref:qgap-amp}) based on the famous
sum-check protocol \cite{ref:Sha92,ref:AroraBarak} and then discuss
how the connection between $\PCP$s and two-player games, which
played a crucial role classically, breaks down in the quantum world,
leading to many new exciting problems. We conclude with a brisk
overview of several points which were not covered in this survey.

\section{Entanglement}\label{sec:entanglement}

In this section we introduce some of the mysterious features of
multipartite entanglement and explain how they affect our basic
understanding of the relationship between states that are
\emph{locally} or \emph{globally} (in)distinguishable --- a
relationship that is at the heart of the classical Cook-Levin
theorem, explored in the next section. 

\subsection{EPR and CAT states} 
\label{sec:multi}

The archetypal example of an entangled state is the
Einstein-Podolsky-Rosen (EPR)
state of a pair of qubits:
\begin{align} 
  \ket{\psi_{EPR}} \EqDef \frac{|00\ra+|11\ra}{\sqrt{2}}
    = \frac{\ket{0}\otimes \ket{0}+\ket{1}\otimes\ket{1}}{\sqrt{2}}.
\end{align} 
It is not hard to show that this state cannot be written as a tensor
product of two single qubit states (try it!). This simple fact
already has interesting consequences. Suppose we measure the left
qubit in the orthonormal basis $\{\ket{0},\ket{1}\}$ (we will not
need to worry much about the formalism of quantum measurements here,
and rather dare to rely on the reader's intuition.) There is a
probability $1/2$ of obtaining the outcome `$1$'; the joint state of
both qubits then collapses to $\ket{11}$ and the right qubit, when
measured, will always yield the outcome `$1$' as well.  Likewise, if
the left qubit is found to be $0$, the right one is also found to be
$0$. The results of the measurements of the left and right qubits
are fully correlated. This full correlation can of course also arise
classically.  The important property is that the EPR state exhibits
similarly strong correlations when measured in \emph{any basis}, not
only the computational one. These correlations, which baffled
Einstein, Podolsky and Rosen themselves, were shown by John
Bell~\cite{Bell:64a} to be stronger than any classical bipartite
system could exhibit. Simplifying Bell's proof, Clauser et
al.~\cite{Clauser:69a} suggested an experiment that could provide
evidence of the inherent \emph{nonlocality} of quantum mechanical
systems. The experiment has since been successfully performed many
times, starting with the work of Aspect in the
1980s~\cite{Aspect82chsh}. 

A system of $2n$  
qubits can be in the product of $n$ EPR pairs.
Though such a state may seem highly entangled, in a sense its
entanglement is not more interesting than that of a single EPR pair;
in particular, the entanglement is \emph{local} in the sense that
each qubit is correlated to only one other qubit. Multiparticle
systems can exhibit far more interesting types of entanglement.  Let
us consider a different generalization of the EPR pair to $n$
qubits, the so-called $n$-qubit CAT states 
\begin{align}
\label{eq:catstate} 
  \ket{\Psi_{CAT}^{\pm}} 
    \EqDef \frac{|0^n\ra\pm|1^n\ra}{\sqrt{2}}
    = \frac{\ket{0}\otimes \cdots 
      \otimes \ket{0}\pm\ket{1}\otimes\cdots\otimes\ket{1}}{\sqrt{2}}.
\end{align}

We would like to argue that the entanglement in the CAT states is
\emph{non-local}, or \emph{global}. To explain this, we need to make
an important detour and introduce the formalism of density matrices,
which provides a mean of representing precisely the information from
a larger quantum state that can be accessed \emph{locally}.

\subsection{Density matrices and global versus local entanglement}
 
Let $\ket{\psi}$ be a state of $n$ particles. Its energy with
respect to a Hamiltonian $H$ can be expressed as 
\begin{align}
  \bra{\psi}H\ket{\psi} 
    = \Tr \big(H \cdot \ket{\psi}\bra{\psi}\big). 
\label{eq:density}
\end{align}
This simple equation gives rise to the definition of a \emph{density
matrix} associated with a pure state: 
\begin{align}
  \label{eq:densitymatrix}
  \rho_{\ket{\psi}}\EqDef \ket{\psi}\bra{\psi} .
\end{align}
For any unit vector $\ket{\psi}$, $\rho_{\ket{\psi}}$ is a rank $1$
positive semidefinite matrix of trace $1$.  More generally, a
density matrix of $n$ qudits, each of dimension $d$, is a $d^n\times
d^n$ positive semidefinite matrix with trace $1$.  Any such $\rho$
may be diagonalized as $\rho = \sum_i \lambda_i \ket{u_i}\bra{u_i}$,
where the $\ket{u_i}$ are orthonormal eigenvectors and $\lambda_i$
the associated eigenvalues, non-negative reals summing to $1$.  The
density matrix has the following useful interpretation: $\rho$
represents a quantum state that is in a \emph{mixture} (a
probability distribution) of being in the pure state $\ket{u_i}$
with probability $\lambda_i$.\footnote{It is important not to
confuse the mixture $\rho$ with the superposition $\sum_i
\sqrt{\lambda_i} \ket{u_i}$; the two are very different states!}
While rank-$1$ density matrices always correspond to pure states,
matrices with higher rank provide a more general way of describing
quantum systems.  Suppose for instance we were only interested in
computing some property of a subset of $k$ particles out of the $n$
particles that are in the pure state $\ket{\psi}$. In general, those
$k$ particles are entangled to the remaining $n-k$ particles, and we
cannot assign to them a pure state.  However, they can be assigned a
density matrix from which the results of all measurements on those
particles can be calculated.  Here is how it can be done. 

Consider some local Hamiltonian $H$ acting solely on the subsystem
$A$ consisting of the $k$ particles: $H = H_A \otimes \Id_B$, where
$B$ denotes the $n-k$ remaining particles. We can compute the energy
\begin{align}
  \bra{\psi}H\ket{\psi} 
    = \Tr \big(H \cdot \ket{\psi}\bra{\psi}\big) 
    = \Tr \big((H_A \otimes \Id_B)\cdot \ket{\psi}\bra{\psi}\big) 
    = \Tr\big(H_A \cdot \Tr_B(\ket{\psi}\bra{\psi})\big).
\label{eq:reduced}
\end{align}
Here we have introduced an important notation, the \emph{partial
trace} operation $\Tr_B$, acting on the density matrix
$\ket{\psi}\bra{\psi}$. Given any matrix $X$ defined on the tensor
product of two Hilbert spaces $\mathcal{H}_A$ and $\mathcal{H}_B$,
its partial trace with respect to $B$ is a matrix on the space of
the particles in $A$, $\mathcal{H}_A$, whose $(i,j)$-th coefficient,
namely, the coefficient on basis states $(\ket{i}_A,\ket{j}_A)$ for
$\mathcal{H}_A$ is defined as
\begin{align}
\label{eq:reduced-2}
  \bra{i}_A\Tr_B(X)\ket{j}_A \,:=\, 
    \sum_k\, (\bra{i}_A\otimes \bra{k}_B) X (\ket{j}_A \otimes\ket{k}_B),
\end{align}
where here $\ket{k}_B$ ranges over an arbitrary orthonormal basis of
$\mathcal{H}_B$ (we leave as a good exercise to the reader to show
that the definition does not depend on the choice of basis for
$\mathcal{H}_B$).  We often refer to the operation of taking a
partial trace on $B$ as \emph{tracing out} the subsystem $B$. If $A$
is empty and $B$ the whole system, we recover the usual definition
of the trace as the sum of the diagonal entries.  A useful property
of the partial trace, which is not difficult to prove, is that it is
commutative: if we divide the system into three subsystems $A,B$ and
$C$ then $\Tr_A(\Tr_B(X)) = \Tr_B(\Tr_A(X))$.  We call the resulting
matrix after we trace out a subsystem $B$, 
$\rho_A=\Tr_B(\ket{\psi}\bra{\psi})$, the \emph{reduced density
matrix} of $\ket{\psi}$ on $A$ (we leave as a second exercise the easy
verification that this is indeed a positive semidefinite matrix with
trace $1$, provided $\ket{\psi}$ is normalized). 

As a further useful exercise, the reader may wish to check that for
a tensor product pure state,
$\ket{\psi}=\ket{\psi_A}\otimes\ket{\psi_B}$, 
$\Tr_B\ket{\psi}\bra{\psi}$ is what one would expect: it is the
density matrix of the state $\ket{\psi}_B$, namely,
$\ket{\psi_B}\bra{\psi_B}$.  In particular, it has rank $1$.  As
soon as $\ket{\psi}$ is entangled, however, its reduced density
matrix will no longer have rank one. One can also verify that the
reduced density on $A$ of 
\begin{align}
\label{eq:schmidt}
  \ket{\psi} = \sum_i \sqrt{\lambda_i} \ket{u_i}_A \ket{v_i}_B, 
\end{align}
where the $\ket{u_i}_A$ and $\ket{v_i}_B$ are orthonormal families, 
is given by 
\begin{align}
\label{eq:schmidt2}
  \rho_A = \sum_i \lambda_i \ket{u_i}\bra{u_i}_A.
\end{align} 

Equipped with the definition of reduced density matrices, let us
return to our discussion of entanglement in the states
$\ket{\Psi_{CAT}^+}$ and $\ket{\Psi_{CAT}^-}$. We claim that locally
these two states look identical.  Indeed, consider any strict subset
$A$ of $k$ qubits, for $k<n$. Then the reduced density matrices of
both states on $A$ are identical.  To see this, take the partial
trace of $\ket{\Psi_{CAT}^+}\bra{\Psi_{CAT}^+}$ and
$\ket{\Psi_{CAT}^-}\bra{\Psi_{CAT}^-}$ over the remaining $n-k$
qubits. Observing that both CAT states~\eqref{eq:catstate} are
written in the form of~\eqref{eq:schmidt}, one can use
Eq.~\eqref{eq:schmidt2} to derive that in both cases $\rho_A =
\frac{1}{2}\ket{0^{k}}\bra{0^{k}} +
\frac{1}{2}\ket{1^{k}}\bra{1^{k}}$.  This is a remarkable feature:
the two CAT states are orthogonal (and hence perfectly
distinguishable), but when tracing out even one of the qubits, they
look exactly the same!  In particular, no local measurement can
distinguish between those states: part of the information carried by
the pair of states is stored in a \emph{global} manner, inaccessible
to local observations. 

Such local indistinguishability of globally distinct states is a
defining feature of multipartite entanglement. The phenomenon,
however, is far richer, and we proceed with the description of a
beautiful example, Kitaev's toric code states \cite{ref:toric},
which demonstrate some of the most counter-intuitive properties of
multipartite entanglement. 

\subsection{The toric code}\label{sec:toric} 

The toric code is defined as the groundspace of a $4$-local
Hamiltonian which acts on a set of $n$ qubits placed on the
\emph{edges} of a $\sqrt{n}\times\sqrt{n}$ two-dimensional grid made
into a torus by identifying opposite boundaries.  To define the
local terms, we first need to introduce the Pauli matrices $Z,X$.
These are $2\times 2$ operators that act on a single qubit. In the
computational basis they are given by
\begin{align*}
  Z &= \begin{pmatrix} 1 & 0 \\ 0 & -1 \end{pmatrix} \ , &
  X &= \begin{pmatrix} 0 & 1 \\ 1 & 0 \end{pmatrix}.
\end{align*}
Both $X$ and $Z$ are Hermitian matrices with eigenvalues $\pm 1$, and they
anti-commute: $ZX = -ZX$.

\begin{figure}
  \begin{center}
    \includegraphics[scale=1]{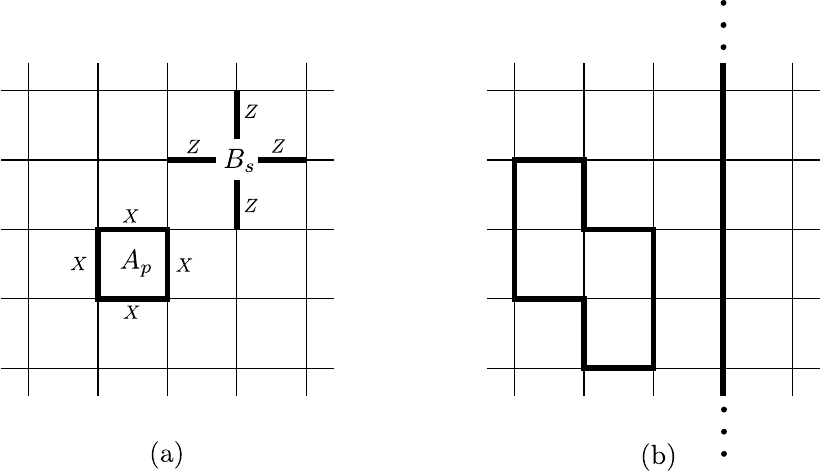}
  \end{center}
  \caption{(a) The plaquette operator $A_p$ and the star operator
  $B_s$. (b) Two loops: a contractible loop (left) and a
  non-contractible loop that wraps around the torus (right).
  \label{fig:toric} }
\end{figure}

The local terms of the Hamiltonian are of two types:
\emph{plaquette} constraints, and \emph{star} constraints. A
plaquette $p$ is a face of the lattice surrounded by $4$ edges (see
\Fig{fig:toric}). To each plaquette $p$ we associate a local term
$A_p$ defined as the product of Pauli $Z$ operators acting on the
qubits that sit on the plaquette's edges.  A star $s$ consists of
the $4$ edges adjacent to a given vertex in the grid (see
\Fig{fig:toric}). To each star $s$ we associate a local term $B_s$
which is the product of Pauli $X$ operators acting on the qubits
sitting on the edges of the star. Hence: 
\begin{align*}
  A_p &\EqDef \prod_{e\in  p} X_e \ , &
  B_s &\EqDef \prod_{e\in s} Z_e \ ,
\end{align*}
where $X_e, Z_e$ denote Pauli matrices acting on the qubit that sits
on the edge $e$. The toric code Hamiltonian is then given by 
\begin{align*}
  H_{toric} \EqDef -\sum_p A_p - \sum_s B_s \ .
\end{align*}

What do the groundstates of $H_{toric}$ look like?  Note first that,
although the $X$ and $Z$ operators anti-commute, a star and
plaquette have either zero or two edges in common, and thus the
corresponding local terms in the Hamiltonian commute: $H_{toric}$ is
a \emph{commuting Hamiltonian}. As a consequence, all local terms
are simultaneously diagonalizable, and any state in the
diagonalizing eigenbasis either fully satisfies or fully violates
any of the constraints.  Although it may seem that this brings us
back to a classical constraint satisfaction scenario, we will soon
see that this is quite far from being the case.  

We now describe a state which is a $1$ eigenstate of all plaquette
and star operators; any such state is necessarily a groundstate.
(This is an example of one state in the four-dimensional groundspace
of the toric code.) Define 
\begin{align}
\label{eq:ground}
  \ket{\Omega} \propto \sum_{loops~s} \ket{s},
\end{align}
where the sum is taken over all computational basis states
associated with $n$-bit strings $s$ such that $s$ is a (generalized)
\emph{loop}: the set of edges in the lattice to which $s$ associates
a $1$ forms a disjoint union of loops, as in \Fig{fig:toric}b, and
we allow loops to wrap around the torus.  To see that $B_s$ leaves
$\ket{\Omega}$ intact, it suffices to show that $B_s$ leaves any
loop $s$ intact. This follows from the fact that the four $Z$
operators forming any $B_s$ intersect a loop in an even number of
positions, hence the $(-1)$ signs due to the action of the $Z$'s 
systematically cancel out.  For the $A_p$, a somewhat more subtle
argument (which we leave as our last exercise) shows that applying
$A_p$ on $\ket{\Omega}$ simply results in a permutation of the order
of summation over loops.  

We point to some remarkable properties of any groundstate of the
toric code.  First we note that applying $X$ operators along any
contractible closed loop, as in the left part of \Fig{fig:toric}(b),
is equivalent to applying all the $A_p$'s of the plaquettes enclosed
by the loop. To see this, use that $X^2=\Id$, so applying $A_p$ on
two adjacent plaquettes cancels the $X$ acting on their
intersection.  Hence not only is the groundstate invariant under the
$A_p$ themselves, but it is also invariant under a much more general
class of operators, comprising any contractible closed loop of
bit-flips ($X$ operators). This is a uniquely quantum phenomenon,
since in the classical world clearly no non-trivial error could
leave a string invariant.  However, notice that this property is not
unique to the toric code --- flipping all the bits in the CAT state 
$\ket{\Psi_{CAT}^+}$ also leaves it invariant.  Now comes a more
surprising property: the relation to the topology of the torus.
Notice that the above argument holds only for a loop that can be
made out of plaquettes, namely, that is contractible. What happens
if one applies a sequence of $X$ along a loop which wraps
\emph{around} the torus (see, for example, the right part of
\Fig{fig:toric}b)?  In that case, the previous argument does not
hold --- but we can argue that the resulting state remains a
groundstate!  Indeed, as before the intersection of such a loop with
any star is even, and thus it commutes with $H_{torus}$; this means
that it keeps the groundspace invariant.  Indeed, one can show that
operators based on non-contractible loops can be used to move
between different groundstates.

The $4$ dimensional groundspace of the toric code can be used as a
quantum error correcting code that encodes $2$ qubits. Its error
correction properties are tightly related to the topological
properties described above; its states are indistinguishable by any
measurement acting on any subset
of qubits of diameter $\orderof{\sqrt{n}}$, and in particular are
globally entangled (we will touch upon this again later, in Section
\ref{sec:NLTS}).\footnote{Though we use the terminology ``global
entanglement'' in a somewhat loose sense, one should note that the
entanglement present in the toric code states is far more complex
than that of the CAT states.  It is sometimes referred to by the
name of \emph{topological order}, and is related to the fact that
these states not only cannot be distinguished locally, but also
cannot be transformed one onto another by local operations alone.
This is a necessary property of a quantum error correcting code, as
otherwise local errors could induce jumps between two states in the
code.\label{footnote:qecc}}

\section{Quantizing CSP results: the local versus global problem} 
\label{sec:quantizing}

Armed with some insight into how complex and beautiful multipartite
entanglement can be, we proceed to explain how its properties affect
our basic understanding of constraint satisfaction problems.

\subsection{Entanglement and local CSPs}
\label{sec:LocalGlobal}

A crucial ingredient in all results related to classical $\CSP$s,
and in particular in the Cook-Levin theorem and the $\PCP$ theorem,
is the ability to verify that a computation is carried out correctly
by performing \emph{local} checks. It is useful to think in this
context of a computation that does \emph{nothing}: all one needs to
verify is that a string has not changed. Classically, this is easily
done by comparing the initial and final strings bit by bit. 

\begin{figure}\label{fig:globalvslocal}
  \begin{center}
    \includegraphics[scale=1]{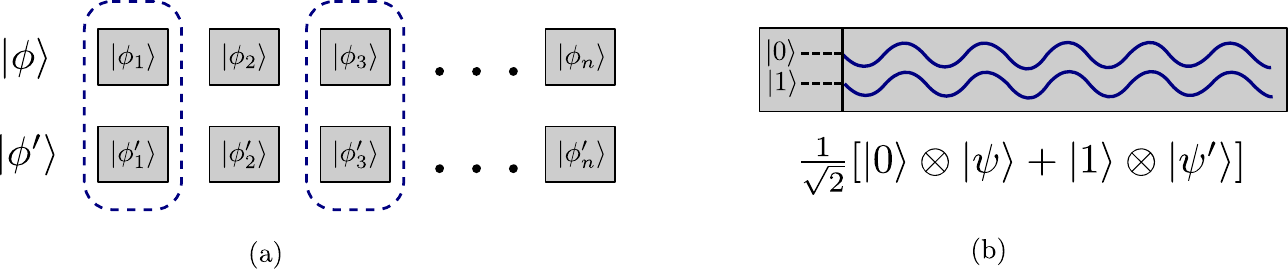}
  \end{center}
  \caption{ (a) Comparing two product states on two registers,
locally.  (b) Two globally entangled states on two separate 
registers cannot be compared locally, but when the two states are in
superposition on the \emph{same} register, comparison can be done 
through a local check of an additional control qubit entangled to
that register. \label{fig:local-checks}}
\end{figure}

How does this extend to the quantum world?  First note that it is
possible to design a local Hamiltonian that ``checks'' if two
\emph{product} states,
$\ket{\phi}=\ket{\phi_1}\otimes\cdots\otimes\ket{\phi_n}$ and
$\ket{\phi'}=\ket{\phi'_1}\otimes\cdots\otimes\ket{\phi'_n}$, are
identical, as follows. For each pair of matching qubits introduce a
local term which projects on the subspace orthogonal to the
symmetric subspace (the space of the two qubits spanned by
$\ket{00}, \ket{01}+\ket{10}, \ket{11}$ (see
Fig.~\ref{fig:local-checks}(a)). The state
$\ket{\phi}\otimes\ket{\phi'}$ will be in the null space of all the 
projections (namely, a groundstate of energy $0$) if and only if the
two states are the same.  

Difficulties arise when trying to design a local Hamiltonian that
checks that two \emph{entangled} states are identical. Recall from
the example of the CAT state that there exists orthogonal states
whose reduced density matrices on any strict subset of the qubits
are identical.  Since the energy of any local Hamiltonian only
depends on those reduced density matrices, no such Hamiltonian can
possibly distinguish between the two states.  This is the crux of
the global-vs-local problem in quantum constraint satisfaction
problems; it limits our ability to locally enforce even such simple
constraints as the identity constraint!  This difficulty will arise
as an important stumbling block in the next subsections, where we
discuss extensions of the Cook-Levin and $\PCP$ theorems to
Hamiltonian complexity. 

\subsection{The quantum Cook-Levin theorem}\label{sec:cooklevin}

In the introduction we presented the local Hamiltonian (LH) problem
as a natural generalization of constraint satisfaction problems to
the quantum domain. We also stated Kitaev's result showing that LH
is for $\QMA$, the class of languages that can be decided in quantum
polynomial time given a quantum state as witness, what $\SAT$ is for
$\NP$: it is a complete problem for the class. The easy direction
states that the $k$-Local Hamiltonian problem is in $\QMA$ for any
constant $k$: verifying that the ground energy of a given LH is
smaller than $a$ can be performed in quantum polynomial time given
the appropriate witness. The idea is quite simple: the witness is an
eigenstate $\ket{\xi}$ of energy lower than $a$, which we are
promised to exist. The verifier can estimate the energy
$E=\sum_{i=1}^m \bra{\xi} H_i \ket{\xi}$ by picking a local term
$H_i$ at random and measuring the witness state $|\xi\ra$ using the
two-outcome projective measurement $\{H_i,\Id-H_i\}$. The
probability that he obtains the outcome associated with $H_i$ is
exactly the average energy $E/m$, which can be estimated to within
inverse polynomial accuracy by repeating the procedure sufficiently
many times (see \Ref{ref:KitaevShenVyalyi} or
\Ref{ref:AharonovNaveh} for more details). 

The other direction, that any language $L\in\QMA$ can be encoded as
an instance of LH, is more interesting. Let us first recall the
proof of the classical Cook-Levin theorem. Suppose $L\in\NP$: there
is a polynomial-time Turing machine $M$ which operates on the input
$x$ and a witness $y$, and is such that there is a $y$ such that
$M(x,y)$ accepts if and only if $x\in L$. To reduce this problem to
a local $\SAT$ formula, introduce a $T \times (T+1)$ table encoding
the history of the computation of $M$ from time $0$ to time $T$. For
each location $1\le i\le T$ on the tape and time $0\le t \le T$
introduce a variable $X_{i,t}$ which is a triple of a tape symbol, a
state of the machine, and a Boolean value indicating whether the
head is at that location or not. The key point is that checking that
the table of $X_{i,t}$ encodes a valid computation can be performed
\emph{locally}. Constraints in the $\SAT$ formula will involve
groups of four variables $X_{t,i-1}, X_{t,i},X_{t,i+1}$ and
$X_{t+1,i}$ and enforce that the state associated with the latter is
a correct result of running $M$ for one step on the state described
by the three former variables. Verifying that the column at time $0$
corresponds to a valid initial state, and the column at time $T$
corresponds to an accepting state, can also be performed locally. 

Suppose now that $L\in \QMA$: there exists a quantum circuit using
two-qubit gates, say $U_1,...,U_T$, which accepts an input $x$ with
probability exponentially close to $1$ provided it is run on the
appropriate witness $\ket{\xi}$ if $x\in L$, and rejects with
probability exponentially close to $1$ if $x \not\in L$, whatever
the witness is.\footnote{Amplification for quantum circuits can be
performed essentially just as for classical circuits, by repeating
sequentially and taking a majority vote. Amplifying to exponentially
small error is not essential here, but it simplifies matters.}
Drawing from the Cook-Levin proof, a natural attempt to encode this
computation into an instance of LH would be to introduce the
sequence of history states of the computation
\begin{align*}
  \ket{\psi_0} &= \ket{x}\otimes\ket{\xi}, \\
  \ket{\psi_1} &= U_1(\ket{x}\otimes\ket{\xi}), \\
               &\vdots \\
  \ket{\psi_T} &= U_T\cdots U_2U_1(\ket{x}\otimes\ket{\xi}),
\end{align*}
place them on $T+1$ adjacent registers, and use local checks to
verify that $\ket{\psi_{t+1}} = U_{t+1}\ket{\psi_t}$ for each $t$.
However, as we have seen in the previous subsection, this approach
is hopeless. Assume for a moment the simple case in which 
$U_{t+1}=\Id$. The presence of entanglement in the states
$\ket{\psi_{t+1}}=U_{t+1}\ket{\psi_{t}}=\ket{\psi_{t}}$ makes it
impossible to use local tests to compare the two different
registers, and verify that the two states are the same, even in the
trivial case when $U_{t+1}=\Id$!

Luckily, entanglement, which is the source of the problem, will also
help us solve it. Consider the state
\begin{align*}
  \ket{\eta}=\frac{1}{\sqrt{2}}\big(\ket{\psi}\otimes\ket{0}
    + \ket{\psi'}\otimes\ket{1} \big),
\end{align*}
in which the states $\ket{\psi}$ and $\ket{\psi'}$ are placed in a
superposition on the \emph{same} register by being entangled to an
additional \emph{control qubit} (see \Fig{fig:local-checks}(b)).
When $\ket{\psi}=\ket{\psi'}$, $\ket{\eta}$ factorizes as
$\ket{\eta} = \ket{\psi}\otimes (\ket{0}+\ket{1})/\sqrt{2}$; the
control qubit becomes unentangled from the first register.
Therefore, if we wish to verify that $\ket{\psi}=\ket{\psi'}$, all
we have to do is make sure that the control qubit is in the state
$(\ket{0}+\ket{1})/\sqrt{2}$.  One can work out that for two general
states $\ket{\psi}, \ket{\psi'}$, if the control qubit in the state
$\ket{\eta}$ is measured in the basis 
$(\ket{0}\pm\ket{1})/\sqrt{2}$, then the probability of getting the
outcome $(\ket{0}+\ket{1})/\sqrt{2}$ is exactly 
$(1+\operatorname{Re}(\bra{\psi}\psi'\rangle))/2$, which is equal to
$1$ if and only if the two states are equal (up to an unimportant
global phase).  This gives a clue as to how to define a local
Hamiltonian that checks that the two states are the same: the
Hamiltonian is defined as the one-qubit projection on the state
$(\ket{0}-\ket{1})/\sqrt{2}$ of the \emph{control qubit} (rather
than on the states themselves!). It is not difficult to verify that
the state $\ket{\eta}$ is in the groundspace of this Hamiltonian
(namely, has energy $0$), if and only if the two states
$\ket{\psi},\ket{\psi'}$ are equal, and otherwise it has a higher
energy. 

Now suppose that we wish to verify not that
$\ket{\psi'}=\ket{\psi}$, but that $\ket{\psi'}=U\ket{\psi}$ for
some local unitary $U$ that acts non-trivially only on $k$ qubits. 
This can be done by generalizing the idea above in a rather
straightforward manner.  In essence, it is simply performing the
same trick, except for first rotating the second state by $U^{-1}$. 
This results in a projection $P$ which acts non-trivially not only
on the control qubit but also on the $k$ qubits on which $U$ acts,
such that the state $\big(\ket{\psi}\otimes\ket{0} +
\ket{\psi'}\otimes\ket{1}\big)/\sqrt{2}$ is in the groundspace of
$P$ if and only if $\ket{\psi'}=U\ket{\psi}$. 

This is the key idea in Kitaev's proof of the quantum Cook-Levin
theorem.\footnote{Kitaev attributes the idea of moving from time
evolution to a time-independent local Hamiltonian to
Feynman~\cite{ref:Feynman1982, ref:Feynman1986}; this construction
is now referred to as the \emph{circuit-to-Hamiltonian} construction 
\cite{ref:adiabatic}.} 
The history of the computation in stored in a \emph{single} register, as
a \emph{superposition} over snapshots describing the state of the
circuit at any given time step. Each such snapshot state is
entangled to the state of an additional clock
register:
\begin{align}
\label{eq:history-state}
  \ket{\eta} = \frac{1}{\sqrt{T+1}}
    \sum_{t=0}^T U_t\cdots U_0
      (\ket{x}\otimes\ket{\xi})\otimes\ket{t} \ .
\end{align}
The Hamiltonian is composed of one term for each time step: the term
associated with time $t$ checks that the state at time $t$ is indeed
equal to $U_t$ applied to the state at time $t-1$.  Each such term
acts non-trivially only on the clock register as well as the qubits
on which $U_t$ acts.\footnote{ Note that here if $t$ is represented
in binary then each term is $\orderof{\log T}$ local. Kitaev also
showed how by representing the clock in unary one could get away
with Hamiltonians that are merely $5$-local.  Further improvements
brought this down to $3$-local~\cite{KempeR033local} and then
$2$-local~\cite{ref:KempeKitaevRegev}.}

To complete the proof of the quantum Cook-Levin theorem one needs to
show that if $x \in L$ then the history
state~\eqref{eq:history-state} has energy less than $a$, but if not
then any state (and not necessarily correct history states) must
have energy substantially larger than $a$.  In the classical
setting, as long as not all constraints are satisfied then at least
one must be violated, so the ``energy'' automatically jumps from $0$
to $1$. In the quantum case this is no longer true, and one has to
perform additional work to prove an inverse polynomial separation
between the two cases. The proof is an elegant collection of
geometrical and algebraic claims related to the analysis of random
walks on the line; the interested reader is referred to
\Ref{ref:AharonovNaveh} for more details.
            

\subsection{Dinur's proof of the $\PCP$ theorem}
\label{sec:dinur}

As shown in the previous section, quantizing the Cook-Levin theorem 
is indeed possible.  Why can't we quantize the proof of the $\PCP$
theorem using similar tricks?  To explain the problem in more
detail, there are two main approaches to the proof of the $\PCP$
theorem to choose from: the original proof~\cite{ref:PCP1,ref:PCP2},
or the recent, more combinatorial one, due to
Dinur~\cite{ref:Dinur-PCP}. Both possibilities are interesting to
explore, and each of them raises issues of a different nature. In
this section we explain the difficulties that arise in quantizing
Dinur's proof, which may a priori seem more accessible due to its
combinatorial nature. At the end of Section~\ref{sec:quantize} we will
briefly return to some of the distinct issues that arise in
quantizing the original proof of the $\PCP$ theorem, which is based
on the use of error correcting codes and procedures for local
testing and decoding.

We proceed to an exposition of Dinur's proof~\cite{ref:Dinur-PCP} of
the $\PCP$ theorem. The proof is based on a general technique called
\emph{gap amplification}. Given an instance $\mathcal{C}$ of some
$k$-local $\CSP$, and an assignment $\sigma$ to the variables of
$\MC$, define the \emph{unsat-value} $\UNSAT_\sigma(\MC)$ of
$\mathcal{C}$ with respect to $\sigma$ as the fraction of clauses of
$\mathcal{C}$ that are not satisfied by $\sigma$. Define the
unsat-value of $\MC$, $\UNSAT(\MC)$, as the minimum over all
$\sigma$ of $\UNSAT_\sigma(\MC)$.  The Cook-Levin theorem states
that it is $\NP$-hard to distinguish between $\UNSAT(\MC)=0$ and
$\UNSAT(\MC)\geq 1/m$, where $m$ is the number of clauses in $\MC$.
Dinur's proof describes a polynomial-time iterative procedure
mapping any instance $\MC$ to an instance $\MC'$ such that, if
$\UNSAT(\MC)=0$ then $\UNSAT(\MC')=0$ as well, but whenever
$\UNSAT(\MC) >0$, then $\UNSAT(\MC') \geq \gap$ for some universal
constant $\gap>0$ (depending only on the class of $\CSP$s from which
$\MC'$ is taken).  Note that {$\gap$} is the relative promise gap
(or simply the promise gap) of the ``amplified'' $\CSP$ $\MC'$.

We focus on $2$-$\CSP$s defined on variables which can be assigned a
constant number of values $d$ (we say that the $\CSP$ is defined
over an alphabet of size $d$), and which have constant degree, where
the degree of a variable is defined as the number of constraints in
which the variable participates. To any such $\CSP$ one can
associate a \emph{constraint graph} $G=(V,E)$ such that each vertex
in $G$ corresponds to a variable, and each edge to a constraint
acting on the variables associated to its two adjacent vertices.
Dinur's gap amplification theorem can be stated as follows. 

\begin{theorem}[Gap amplification -- adapted from Theorem~1.5
  in~\Ref{ref:Dinur-PCP}]
\label{thm:Dinur}
  For any alphabet size $d$, there exist constants $0<\gap<1$ and
  $W>1$, together with an efficient algorithm that takes a constraint
  graph $G=(V,E)$ with alphabet size $d$, and transforms it to
  another constraint graph $G'=(V',E')$ with universal alphabet
  size $d_0$ such that the following holds: 
  \begin{enumerate}
    \item $|E'|\le W|E|$ and $|V'|\le W|V|$,
    \item (completeness) if $\UNSAT(G)=0$ then $\UNSAT(G')=0$,
    \item (soundness) if $\UNSAT(G)>0$ then 
      $\UNSAT(G')\ge \min\big(2\cdot\UNSAT(G), \gap\big)$.
  \end{enumerate}
\end{theorem}

Starting from a constraint graph associated to an instance of a
$\NP$-hard $\CSP$ (such as $3$-coloring), the $\PCP$ theorem follows
by applying the above theorem logarithmically many times.

Let us try to na\"{\i}vely prove Theorem \ref{thm:Dinur}. Given an
instance $\MC$, there is a simple way to perform gap amplification:
given an integer $t$, construct a new instance $\MC_t$ which has the
same variables as $\MC$ but $m^t$ constraints corresponding to all
possible conjunctions of $t$ constraints drawn from $\MC$. It is
easy to see that $\UNSAT(\MC_t) \geq 1-(1-\UNSAT(\MC))^t$.
Unfortunately, this construction has a major drawback: in order to
reach a constant promise gap, we need to choose $t=\Omega(m)$.
Verifying a random constraint requires querying a constant fraction
of the variables, and is therefore useless for a $\PCP$ proof.

To overcome this problem, Dinur starts from the above idea but
breaks it into small steps; at each step she performs an
amplification by a constant $t$, which is then followed by a
regularizing step that restores the system's locality without
substantially damaging the amplification of the gap. This approach,
however, raises another problem: since at each step the number of
constraints grows like $m\to m^t$, and since the final system can be
at most polynomially large, then even for a constant $t$, we can
only perform a constant number of iterations -- which will result only
in a constant amplification. The key idea in Dinur's construction is
to use expander graphs to overcome this difficulty. Expander graphs
are low-degree graphs with the property that a random walk on the
graph is rapidly mixing, and quickly reaches the uniform
distribution over all vertices. This property suggests a more
efficient way to perform gap amplification: instead of including
\emph{all} possible subsets of $t$ constraints in $\MC_t$, only
include constraints (edges) obtained as the conjunction of $t$
constraints that form a path in the constraint graph.  We call such
paths \emph{$t$-walks}. The property of rapid mixing ensures that
constraints constructed in this way are sufficiently
well-distributed, so that the promise gap is amplified almost as
much as in the previous construction. However, the number of new
constraints is significantly smaller: for an expander graph with a
constant degree $D$, the number of constraints is at most $m\cdot
D^t$ instead of $m^t$. We may thus iterate this process
logarithmically many times, resulting in the desired amplification
to a constant promise gap.

This basic idea leaves us with two difficulties on the way to the
proof of Theorem~\ref{thm:Dinur}. First, the constraint graph should
be an expander.  Second, the above suggestion increases the locality
of the system and replaces $2$-local terms with $t$-local ones.
Dinur's proof uses additional ideas to handle these issues.
Altogether, her construction has three main steps:
\begin{description}
  \item [Preprocessing.]
    This step turns the constraint graph into an expander. The
    crucial part is called \emph{degree reduction}, in which the
    degree of
    every vertex is reduced to $D_1+1$ for some constant $D_1$. This
    is achieved by replacing every vertex of degree $\ell>D_1+1$
    with a ``cloud'' of $\ell$ vertices, one for each edge,
    connected by an expander of degree $D_1$: their degree becomes
    $D_1+1$, where the extra $1$ is due to the original, external,
    edge. Consistency between the new vertices is enforced by
    placing identity constraints on the edges of the expander. 
    
  \item [Gap amplification.]
    This step uses the idea outlined above to amplify the gap using
    $t$-walks on the constraint graph. This most naturally leads to
    a constraint \emph{hypergraph}, and Dinur's proof introduces
    additional ingredients (that we will not describe) to turn it
    back into a graph; once again, this part involves adding
    consistency constraints.
    
  \item [Alphabet reduction.]
    The previous step leaves us with a constraint graph with an
    amplified promise gap, but with very large alphabet size (in
    fact, it is doubly exponential in $t$).  To reduce it, Dinur
    uses a construction known as \emph{assignment
    tester}~\cite{ref:AssignmentTesters}; it is also known by the
    name of \emph{$\PCP$ of proximity}~\cite{ref:PCPP}. This
    transformation uses error correcting codes to transform any set
    of constraints into a larger set of constraints on a larger
    number of variables, but such that the variables now range over
    an alphabet of constant size $d_0$ which is the universal
    alphabet size from Theorem \ref{thm:Dinur}.  Constraints are
    added to check that the assignment is really a word in the code.
    The code must thus be \emph{locally testable}, meaning that if
    the assignment is far from satisfying it will necessarily
    violate many constraints. 
\end{description}
We remark that the first and last steps reduce the promise gap by
constant factors (which is of course something we want to avoid), 
but by choosing a sufficiently large (yet constant) $t$ in the
second step, one can show that the overall promise gap is still
amplified. 

\subsection{Trying to quantize Dinur's construction} 
\label{sec:quantize}

Consider the natural quantum analogue of a constraint graph
$G=(V,E)$: a $2$-local Hamiltonian acting on $d$-dimensional
particles placed on the vertices of $G$, where to every edge $e\in
E$ we associate a $2$-local projection $H_e$ acting on the two
adjacent particles. The result is a $2$-local Hamiltonian $H_G =
\sum_{e\in E} H_e$, which we will refer to as a quantum constraint
graph. The unsat-value of a classical assignment becomes the quantum
unsat-value of a state $\ket{\psi}$ with respect to $H_G$: 
\begin{align}
\label{eq:qunsat}
  \QUNSAT_\psi(H_G)\EqDef
    \frac{\bra{\psi}H\ket{\psi}}{m} = 
      \frac{1}{m}\sum_{i=1}^m\bra{\psi}H_i\ket{\psi}. 
\end{align} 
This is just the average energy of the local terms $H_i$.  The
unsat-value of $H_G$ is then the minimum quantum unsat-value over
all states. By definition, it is reached by the groundstate
$\ket{\gs}$ of $H_G$, and therefore $\QUNSAT(H_G)=\Egs/m$, where
$\Egs$ is the groundstate energy. 

With this analogy, one is tempted to find quantum equivalents of the
three steps in Dinur's proof. In light of
Section~\ref{sec:LocalGlobal} this seems a nontrivial task. Take,
for example, the degree reduction part in the preprocessing step of
Dinur's proof. Classically, one replaces a vertex with degree $D_1$
by a cloud of $D_1$ vertices with identity tests in-between them.
How can we achieve an analogous construction in the quantum world? 
There seem to be two problems here. The first is the issue of
entanglement.  The particle we would like to ``copy'' is potentially
entangled with additional vertices, and its state is unknown; it is
unclear how to map the state of the particle and the rest of the
system to a state in which there are more ``identical copies'' of
that particle. Moreover, even if such a map could be defined, it is
unclear how local Hamiltonians could be used to check that the
resulting state has the required properties, given the impossibility
of local consistency checking in the quantum setting (as described
in Section~\ref{sec:LocalGlobal}).  Similar difficulties arise in
the other steps of the classical construction: in each step new
variables are introduced, and consistency checks added; quantizing
each such check presents an additional challenge. 

There is at least one non-trivial step in Dinur's proof which we do
know how to quantize, as it avoids the above-mentioned difficulty of
``consistency checking of newly added variables''. 
This is the gap amplification by $t$-walks on expanders, in which
$2$-local constraints are replaced by conjunctions of $t$-local 
constraints along paths of the expander graph.  In the quantum
analogue~\cite{ref:qgap-amp}, the $2$-local terms are replaced by
new terms (alternatively, quantum constraints) defined on $t$-walks.
The new quantum constraint on a given $t$-walk is the conjunction of
all the old $2$-local constraints along the $t$-walk; namely, its
null space is defined as the intersection of all the null spaces of
the old constraints. The proof that this indeed amplifies the gap 
in the quantum setting is non-trivial, and requires much technical
work. Unfortunately, that it can be done still does not provide a
hint as to how to overcome the aforementioned difficulty of adding
new variables and checking consistency. 

One is tempted to try and resolve this issue using the  
circuit-to-Hamiltonian construction,
which was useful in overcoming
the consistency check problem in the quantum Cook-Levin
proof (see \Sec{sec:cooklevin}). So far, all attempts we are aware
of to follow this path have led to an unmanageable reduction of the
promise gap. Another possibility would be to use quantum
gadgets~\cite{ref:KempeKitaevRegev, ref:TerhalOliveira,
ref:Terhal-Gadgets}.  Such gadgets allow moving from $k$-local to
$2$-local Hamiltonians on a larger system of particles, while not
changing the groundstate too much (and preserving the existence of a
$\Gap=\Omega(m)$ absolute promise gap).  One might hope to apply the
gadgets to the construction of~\cite{ref:qgap-amp} mentioned above
in order to achieve a quantum analogue of the full gap amplification
step in Dinur's proof. However, while the best constructions so far
\cite{ref:Terhal-Gadgets} approximately preserve the ground energy,
they also increase the number of terms and their norm by a constant
factor; this results, yet again, in an unmanageable decrease of the
relative promise gap.\footnote{The gap decreases like $\gap \mapsto
\gap^{\poly(k)}$; this is a simple though implicit corollary from
Sec.~III of \Ref{ref:Terhal-Gadgets} (arXiv version).}

The difficulties described above have sometimes been
attributed~\cite{ref:AaronsonQPCP} to the no-cloning
theorem~\cite{ref:no-cloning}, which asserts that unknown quantum
states cannot be copied. However, no-cloning only applies to
\emph{unitary} transformations; there is no reason to require that a
quantum $\PCP$ transformation mapping the groundstate of $H$ to that
of an $H'$ with amplified gap be a quantumly implementable map
(indeed, the quantum map is from $H$ to $H'$, but its action on the
groundstate need not be quantumly implementable by itself).  It
seems that the central issue lies in the combination of the
difficulty of locally copying the state of a particle that might be
entangled to the rest of the system, with that of locally comparing
two states that are supposed to be equal.  

We conclude this section with a remark about the $\qPCP$ conjecture
and quantum error correcting codes (QECC). In classical $\PCP$
proofs, one often uses some kind of error correcting code to encode
the initial configuration space \emph{and} the initial constraints.
The encoded $\CSP$ will have two types of local constraints; one to
check that we are inside the code, and a second to check that the
original constraints are satisfied. Quantumly, however, this seems
impossible.  Even though we do have quantum codes which can be
specified by local constraints (e.g., the toric code), it is
\emph{by definition} impossible for local tests to distinguish
between codewords encoding different (orthogonal) states.  This
indistinguishability is necessary to protect the information from
being destroyed by local interactions with the environment. 
Indeed, if this were not the case then the environment could
effectively acquire information about the encoded state by locally
measuring the codeword, thus potentially destroying any quantum
superposition present in the state through a single local operation.
Thus, if the groundstate arising from the $\qPCP$ reduction comes
from such a quantum error correcting code, then any $k$-tuple of the
qubits will not reveal any information on the encoded state. As a
consequence it is unclear how the original local constraints could
be verified; to the least this cannot be done by locally decoding
each of the qubits on which the original constraint acted.  

\subsection{Brand{\~a}o-Harrow's limitations on qPCP} 
\label{sec:BH}

Is it possible to put the intuitive obstructions discussed in the
previous section on formal grounds, thereby deriving a refutation,
or at least strong limitations on the form that $\qPCP$ can take?
There have been several recent attempts along these
lines~\cite{ref:nogo, ref:BH,ref:DoritLior2}.  We present here the
strongest result so far, due to Brand{\~a}o and Harrow~\cite{ref:BH}
(we will return to the other two results in Section \ref{sec:NLTS}).
The result of Brand{\~a}o and Harrow imposes limitations both on the
form of the local Hamiltonians that could possibly be the outcome of
quantum $\qPCP$ reductions, as well as on the reductions themselves.

The general approach of \Ref{ref:BH} is to identify conditions under
which the approximation problem associated with the $\qPCP$
conjecture is inside $\NP$, and therefore cannot be $\QMA$-hard
(assuming $\NP\neq \QMA$).  More precisely, they identify specific
parameters of a 2-local Hamiltonian (as well as its groundstate)
such that when the parameters lie in a certain range there is
guaranteed to exist a product state whose average energy is within
the relative promise gap $\gap$ of the ground energy of the
Hamiltonian.  Such a product state can then serve as a classical
witness, putting the approximation problem in $\NP$. The proof is
based on information-theoretic techniques 
and is inspired by methods due to Raghavendra and
Tan~\cite{ref:Raghavendra}, which they introduced to prove the fast
convergence of the Lasserre/Parrilo hierarchy of semidefinite
programs for certain $\CSP$s.  

We first state the theorem formally and then discuss the dependence
of the error term on various parameters of the Hamiltonian. 
\begin{theorem}[Groundstate approximation by a product state
  (adapted from \Ref{ref:BH})] \label{thm:gs-approx} Let $H$ be a
  $2$-local Hamiltonian defined on $n$ particles of dimension $d$,
  whose underlying interaction graph (the graph
  whose vertices represent the particles, and which has an edge for
  every local Hamiltonian term) has degree $D$. Then for every state
  $\ket{\psi}$, integer $\bran>0$, and partition of the vertices
  into subsets $\{X_i\}_{k=1}^{n/\bran}$ each composed of $\bran$
  particles, there is a product state $\ket{\phi} =
  \ket{\phi_1,\ldots,\phi_{n/\bran}}$ (with $\ket{\phi_i}$ a state
  on the particles associated with $X_i$) such that the average
  energies $\QUNSAT_\psi(H)$ and $\QUNSAT_\phi(H)$ of
  $\ket{\psi},\ket{\phi}$ with respect to $H$ (as defined
  in~\eqref{eq:qunsat}) satisfy
  \begin{align}
    \label{eq:BH}
    \big|\QUNSAT_\psi(H)
 - \QUNSAT_\phi(H)\big|
      \le  W\cdot\left(\frac{d^6 \Av_i\Phi_G(X_i)}{D}
        \cdot\frac{ \Av_i S_\psi(X_i)}{\bran}\right)^{1/8} 
      \EqDef \eta \ ,
  \end{align}
  where $W$ is a universal constant, $\Av_i$ denotes averaging with
  respect to the subsets $\{X_i\}$, $\Phi_G(X_i)$ is the \emph{edge
  expansion} of $X_i$, and $S_\psi(X_i)$ the von Neumann entropy of
  the reduced density matrix of $\ket{\psi}$ on $X_i$ (see below for
  precise definitions).
\end{theorem}

Let us consider the various parameters on which the error term
$\eta$ depends.

\begin{description}
 \item [Graph degree $D$:] 
   As one can see, the higher the degree $D$, the smaller the error
   $\eta$.  This is the most interesting aspect of the result, as it
   should be contrasted with the fact that in classical $\PCP$s the
   degree can be made arbitrarily large; we return to this point in
   more detail below. It is a manifestation of an important property
   of multipartite entanglement known as \emph{monogamy of
   entanglement} \cite{ref:monogamy}.  Intuitively, monogamy states
   that a particle cannot be simultaneously highly entangled with
   many other particles: the more particles it is entangled to, the
   less it is entangled with each particle. Therefore, in the
   groundstate of a high degree system, the particles are not
   expected to be highly entangled with each other on average, and a
   product state provides a good approximation. 

  \item [Average Expansion $\Av_i \Phi_G(X_i)$:]
	The \emph{edge expansion} $\Phi(X_i)$ of a set of vertices $X_i$
    is the ratio between the number of edges that connect $X_i$ with
    $V\setminus X_i$ and the total number of edges that have at
    least one of their vertices in $X_i$.  Seemingly, the dependence
    of the error on the average expansion $\Phi_G$ is
    non-surprising; we expect the approximation problem to be harder
    for good expander graphs, and so it is expected for $\eta$ to
    increase as $\Av_i \Phi_G(X_i)$ increases. Yet, using the bound
    $\Phi_G(X_i) \le\Phi_G\le 1/2 - \Theta(D^{-1/2})$
    \cite{ref:Expanders}, where $\Phi_G$ is the edge expansion of
    the graph, we find $D^{-1} =
    \mathcal{O}\big((1/2-\Phi_G)^{2}\big)$ and hence $\eta =
    \mathcal{O}\Big[(1/2-\Phi_G)^{1/4}\Big]$ (where we also used
    $\Phi_G\leq 1$). Therefore, very good expanders, whose edge
    expansion approaches the maximum value $1/2$, are not candidates
    for $\QMA$-hard instances of the approximation problem.  Note
    that for the expansion to be close to its optimum value $1/2$,
    by the above bound the graph should have very high degree, so
    this can actually be viewed as a consequence of the previous
    item. Given the role of expanders in Dinur's proof of the
    classical $\PCP$,
		 this again can be interpreted as strong evidence against
    $\qPCP$ .\footnote{A similar phenomenon showing that
    very good expanders are not candidates for $\QMA$-hard instances
    was independently discovered by Aharonov and Eldar 
    \cite{ref:DoritLior2} in the different context of commuting
    Hamiltonians on hypergraphs; this is discussed in Section
    \ref{sec:commutingbad}.}

  \item [Average entanglement:]
	The von Neumann entropy of a density matrix is the usual Shannon
    entropy of its eigenvalues and is defined as $S_\psi(X_i) = -\Tr
    \rho_i\log(\rho_i)$, where $\rho_i$ is the reduced density
    matrix of $\ket{\psi}$ on $X_i$ (see
    \Sec{sec:multi}).\footnote{The von Neumann entropy is also known
    as the \emph{entanglement entropy}, as it measures the amount of
    entanglement between $X_i$ and the rest of the system; in the
    case of a product state $\rho_i$ has rank $1$ and the entropy is
    $0$.} Note that here $S_\psi(X_i)/\bran$ is at most $\log(d)$,
    since there are $\bran$ particles, each of dimension $d$, in
    $X_i$.\footnote{Just like the Shannon entropy, the von Neumann
    entropy is bounded by the logarithm of the dimension.} For a
    $\qPCP$ to be possible, the average entanglement must thus be
    $\Omega(\bran)$, proving that not only must the state be highly
    entangled, but subsets of particles should carry entropy of the
    order of the number of particles they consist of; speaking in
    geometrical terms, the average entanglement entropy should be of
    the order of the \emph{volume} of $X_i$.\footnote{The reader
    might be reminded of a physical phenomenon known as \emph{area
    law} \cite{ref:AL-rev}, by which the groundstate entanglement
    entropy of a region scales like its ``area'' (namely, the number
    of terms connecting particles in and out of it) rather than its
    ``volume'' (its number of particles). This phenomenon is not
    relevant here: if the $X_i$ have constant expansion, ``area''
    and ``volume'' scale similarly, whereas if the expansion is bad,
    a trivial state exists due to a much simpler argument
    (essentially by disconnecting the subsets $X_i$ from one from
    another); see Section \ref{sec:commutingbad}.} 
\end{description}    

The strength of Brand{\~a}o and Harrow's result \cite{ref:BH} lies
not only in the set of Hamiltonians ruled out as possible hard
instances of LH, but also in excluding a very large set of possible
\emph{mappings} that one may want to construct in order to prove the
$\qPCP$ conjecture.  Indeed, first note that even if all parameters
appearing as numerators in \eqref{eq:BH} take their maximal value,
we have $\eta\le
W\cdot(\frac{\log(d)d^6}{2D})^{\frac{1}{8}}<\frac{Wd}{D^{1/8}}$. We
can assume without loss of generality that $d/D^{1/8}<1/2$, since we
can always increase $D$ by adding a constant number of edges to each
vertex with trivial projections, whose energy is always $0$. This
would decrease the promise gap by at most a constant factor. Let us
now assume that there exists an efficient mapping that takes a
$2$-local Hamiltonian with particle dimension $d$, and whose 
underlying interaction graph has degree $D$, and transforms it into
a new instance of LH with particle dimension $d^2$, and whose
underlying interaction graph $G'$ has degree $D^2$, without
\emph{decreasing} the promise gap.\footnote{Presumably, the goal of
such a mapping, which would perform some kind of squaring of the
constraint graph, would be to \emph{increase} the promise gap from
$\gap=1-(1-\gap)$ to $\gap'=1-(1-\gap)^2\approx 2\gap$.} Speaking
loosely, such a reduction will take $\eta\to\frac{1}{2}\eta$, and
applying it $t=\orderof{\log (\gamma^{-1})}$ times, where $\gap$ is
the initial relative promise gap, we will get $\eta<\gap$, which
would place the problem inside $\NP$. Notice that we could also
apply this reduction to a general $k$-local Hamiltonian, by first
turning it into a $2$-local Hamiltonian using the gadgets machinery
of \Ref{ref:Terhal-Gadgets}; this would only reduce the initial
relative promise gap by a constant factor. We therefore arrive at
the following surprising corollary: 

\begin{corol}[Adapted from \cite{ref:BH}]\label{coro:BH}
  The existence of an efficient classical reduction that 
  takes a quantum constraint graph with particle dimension $d$
  and degree $D$ to a new quantum 
  constraint graph with particle
  dimension at most $d^2$ and degree at least $D^2$ without 
  decreasing the promise gap is
  incompatible with the $\qPCP$ conjecture being true.
\end{corol}

We note that in the classical world, such a mapping trivially
exists. Indeed a rather simple construction based on the idea of
parallel repetition would do the trick (see Proposition~4 in
\Ref{ref:BHfull} for an explicit description). Moreover, these
constructions are the bread-and-butter of classical $\PCP$ proofs.
In fact, a closer inspection of Dinur's $\PCP$ proof reveals that
the second step in her gap amplification theorem gives exactly such
a reduction!\footnote{It does more in fact --- it also amplifies the
promise gap.} We arrive at a paradoxical situation in which
quantizing the second step of Dinur's proof would imply that $\qPCP$
does \emph{not} hold.  We conclude that one cannot prove the $\qPCP$
conjecture by directly mimicking Dinur's classical proof. 

\section{The NLTS conjecture: room temperature entanglement} 
\label{sec:NLTS}

Given the difficulties outlined above in proving \emph{or}
disproving the $\qPCP$ conjecture, Hastings suggested a seemingly
easier conjecture~\cite{ref:NLTS1,ref:NLTS2}, the \emph{no
low-energy trivial states} (NLTS) conjecture.  This conjecture is
implied by the $\qPCP$ conjecture, but the other direction is not
known.  The $\NLTS$ conjecture is of interest on its own, as it
captures elegantly the essence of the physical intuition as to why
the $\qPCP$ might not hold. To state it, Hastings chooses a
characterization of multipartite entanglement through the notion of
\emph{non-trivial states}.  A state $\ket{\psi}$ is said to be
trivial if it is the output of a constant depth quantum circuit
applied to the input state {$\ket{0^n}$}.\footnote{A slightly more
general definition of a trivial state, used in \Ref{ref:NLTS2},
defines it as a state $\ket{\psi}$ that can be \emph{approximated}
(to some prescribed accuracy $\epsilon>0$) by a state generated by a
constant depth quantum circuit (we might refer to this as
\emph{approximately trivial}).  When discussing the general $\NLTS$
and $\qPCP$ conjectures, the two definitions essentially amount to
the same.  This is because these two conjectures already contain the
notion of approximation.  In particular, by slightly increasing the
gap in any one of these conjectures, one can switch to talking about
exactly trivial states rather than approximately trivial ones.}
\begin{conj}[NLTS conjecture]
\label{con:NLTS} 
  There exists a universal constant $c>0$, an integer $k$ and a family of $k$-local
  Hamiltonians $\{H^{(n)}\}_{n=1}^\infty$
  such that for any $n$, $H^{(n)}$ acts on $n$ particles, and
  \emph{all} states of average energy less than $c$ above the 
average ground energy with respect to
  $H^{(n)}$ are non-trivial. 
\end{conj} 

Let us see why the $\NLTS$ conjecture is implied by the $\qPCP$
conjecture. Suppose that the $\qPCP$ conjecture holds with relative
promise gap $\gap >0$. We claim that the family of Hamiltonians
produced by the $\qPCP$ reduction (or an infinite subfamily of it)
satisfies the $\NLTS$ requirements for $c=\gap$. Indeed, consider a
Hamiltonian which has a trivial low-energy state of average energy
below $\epsilon_0+\gap$, where $\epsilon_0=\Egs/m$ is the ground
energy averaged over the number of constraints. Then the circuit
generating the state can serve as a classical witness from which one
can efficiently compute the energy classically (since the circuit
generating the state is of constant depth) and verify that the
Hamiltonian indeed has average ground energy less than
$\epsilon_0+\gap$.  This would place the approximation problem
inside $\NP$, contradicting the $\qPCP$ conjecture as long as
$\QMA\neq\NP$. 

What about the other direction?  As mentioned above, it seems that the
$\NLTS$ conjecture is weaker than $\qPCP$ and does not necessarily
imply it.  This makes the $\NLTS$ conjecture an interesting target
to attack. Probably easier and more accessible than the $\qPCP$
conjecture, its resolution is likely to shed some
light on the $\qPCP$ conjecture and the Hamiltonians required for
its proof, if such a proof exists.

What makes the $\NLTS$ conjecture an interesting milestone is the
choice of states that it restricts attention to: the class of
non-trivial states.  This class seems to elegantly capture some very
interesting characteristics of \emph{global entanglement}, as 
discussed in Section~\ref{sec:entanglement}. Formally, we say that a state
$\ket{\psi}$ is \emph{globally entangled} if there exists a state
$\ket{\psi'}$ orthogonal to it such that $\bra{\psi}O\ket{\psi} =
\bra{\psi'}O\ket{\psi'}$ for every local operator $O$.  We have seen
in Section~\ref{sec:entanglement} that both the CAT
states~\eqref{eq:catstate} and the toric code states are globally
entangled.\footnote{As well as any state in a non-trivial quantum
error correcting code; see footnote~\ref{footnote:qecc}.} It turns
out that any globally entangled state is non-trivial
\cite{ref:HBV06}: suppose that $\ket{\psi}$ is globally entangled,
so that it is indistinguishable locally from some $\ket{\psi'}$, and
suppose for contradiction that $\ket{\psi}=U\ket{0^{n}}$ with $U$ a
constant depth quantum circuit. Let $O$ be a local operator.  Then
$UOU^{-1}$ is also a local operator (as can be seen by by tracking
the ``light cone'' of $O$ through the circuit $U$). Hence $
\bra{\psi'}UOU^{-1}\ket{\psi'} = \bra{\psi}UOU^{-1}\ket{\psi} =
\bra{0^{n}}O\ket{0^{n}}\ .$ Taking for $O$ the projection on
$\ket{0}$ of any qubit, we find a state $U^{-1}\ket{\psi'}$ that is
equal to $\ket{0}$ in all qubits. Applying $U$, we get
$\ket{\psi'}=U\ket{0^n}=\ket{\psi}$, a contradiction.  This 
observation helps motivate the study of non-trivial states.

Most of the recent results on $\qPCP$, including the main result of
Brand{\~a}o and Harrow~\cite{ref:BH} reviewed in
Section~\ref{sec:BH}, can be interpreted directly as progress on the
$\NLTS$ conjecture.  For example, the main result of~\cite{ref:BH}
can be viewed as ruling out Hamiltonians whose degree grows
asymptotically with $n$ from being good candidates for proving the
$\NLTS$ conjecture.  In this section we survey additional recent
progress on this conjecture (both negative and positive) through
several results~\cite{ref:BV,ref:NLTS1, 
ref:DoritLior1,ref:DoritLior2,ref:schuch,ref:Hastings2D}.  All of
those results apply to a special subclass of local Hamiltonians
called \emph{commuting Hamiltonians}, in which the terms of the
Hamiltonian are required to mutually commute. As we have seen from
the toric code example (\Sec{sec:toric}), important insights can
already be gained by considering this special case; moreover, the
commuting restriction imposes additional structure that makes the
mathematics involved significantly simpler. 

The remainder of this section is organized as follows.  We start in
Section~\ref{sec:nontrivial} with some simple observations regarding
conditions on Hamiltonians that could possibly serve as good
candidates for the $\NLTS$ conjecture: such Hamiltonians must be
good expanders.\footnote{This also answers the natural question of
why the toric code isn't a good candidate for the $\NLTS$
conjecture: its underlying interaction graph is not expanding.} In
the following section we provide background on commuting
Hamiltonians, and survey several results that provide further
restrictions on the Hamiltonians for which the $\NLTS$ conjecture
may hold.  Lastly, in Section~\ref{sec:freedmanhastings} we survey a
recent construction due to Freedman and Hastings \cite{ref:NLTS2}
that suggests a possible route towards a positive resolution of the
conjecture.

\subsection{Some simple observations} 
\label{sec:nontrivial}

As we have seen, the toric code already provides a family of
Hamiltonians whose groundstates are globally entangled, hence
non-trivial.  Why doesn't the toric code satisfy the conditions of
the $\NLTS$ conjecture? It turns out that as soon as one considers
states with energy slightly above the ground energy, one finds
trivial states. This is a simple consequence of the fact that the
toric code is embedded on a two-dimensional lattice (the argument is
the same for any constant dimension).  To see this, partition the
lattice into squares of size $\ell \times \ell$, and consider the LH
obtained after removing all local terms that act on the boundary
between two different squares. This results in a union of
disconnected local Hamiltonians; finding a groundstate on each
square and taking their tensor product will yield a groundstate of
the new LH. Thus for any constant $\ell$, the new Hamiltonian has a
trivial groundstate. However, notice that the energy of this state
with respect to the original Hamiltonian is not large: the
difference is at most the energy of the local terms we threw away,
which in terms of average energy is $\orderof{1/\ell}$.\footnote{The
toric code can also be defined in
$4D$~\cite{ref:DennisKitaevPreskillLandhal}. There is a notion
according to which the $4D$ construction does have robust
entanglement at room temperature~\cite{ref:nonzeroTQO}, but still it
does not satisfy $\NLTS$, since the same argument against the $2D$
toric code being an $\NLTS$ applies.} 

More generally, for a Hamiltonian to be $\NLTS$ its underlying
interaction graph must be highly connected, so that small sections
cannot be isolated by removing a small number of edges --- a
condition directly associated with the notion of expansion.  Can we
find $\NLTS$ Hamiltonians among $2$-local Hamiltonians defined on
expanders?  This brings us back to the discussion of \Ref{ref:BH} in
Section~\ref{sec:BH}. Re-interpreted in terms of $\NLTS$, the result
states that any family of $2$-local Hamiltonians defined either on
good enough expanders or graphs with sufficiently large degree
cannot satisfy the $\NLTS$ condition.  What about $2$-local
Hamiltonians defined on graphs whose expansion properties lie
in-between the two extreme cases of very good expanders (those with
asymptotically growing degree) and low-dimensional lattices? To the
best of our knowledge, nothing is known about Hamiltonians in that
range so far.

A natural direction to pursue is to turn to the study of $k$-local
Hamiltonians, for $k>2$. The choice of $k$ might seem unimportant in
our context: just as in the case of the $\qPCP$ conjecture, using
the gadgets of \cite{ref:Terhal-Gadgets} one can reduce any
{$\NLTS$} Hamiltonian with {$k>2$} to an {$\NLTS$} Hamiltonian with
$k=2$. However, these gadgets do not preserve the geometry of the
graph (such as expansion), nor do they preserve the commutation
relations. Hence, when studying the $\qPCP$ conjecture or the
$\NLTS$ conjecture for \emph{restricted} subclasses of Hamiltonians
(e.g. highly expanding graphs or commuting Hamiltonians), the
parameter $k$ might in fact play an important role.  In the
following, we present various results that apply to some specific
subclasses of $k$-local Hamiltonians with $k>2$. To the best of our
knowledge all results in this context apply only to commuting local
Hamiltonians.

\subsection{Further limitations on NLTS: the commuting case}
\label{sec:commutingbad} 

\begin{figure}
  \begin{center}
    \includegraphics[scale=1]{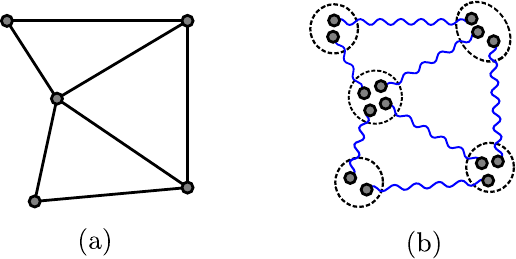}
  \end{center}
  \caption{\label{fig:BV} (a) The original $2$-local Hamiltonian; 
    (b) After restricting each particle to one of its subspaces in the 
    direct sum of the Bravyi-Vyalyi structure lemma, the Hamiltonian
    becomes a disjoint collection of pairwise interactions, whose
    groundstate is a tensor product of states of pairs of subparticles.}
\end{figure}

In the remainder of this section we restrict our attention to
commuting local Hamiltonians ($\CLH$s), in which all terms of the
Hamiltonian mutually commute.  The key statement that makes the
situation in the case of $\CLH$s simpler is a fundamental structural
lemma due to Bravyi and Vyalyi (BV)~\cite{ref:BV}, which we now
describe.  Although the original BV lemma was proven in the context
of commuting $2$-local Hamiltonians, we will then see how it can
also be used to analyze $k$-local Hamiltonians. 

Suppose thus that $H$ is $2$-local and let $G=(V,E)$ be the
associated constraint graph. Consider a $d$-dimensional particle
associated with a vertex $v$ of degree $D$, so that there are $D$
local terms that act on the particle.  BV show that the commuting
condition implies that these terms can essentially be viewed as
acting on $D$ distinct, \emph{disjoint} ``sub-particles''. Formally,
they show that the Hilbert space $\mathcal{W}^{(v)}$ of the particle
at $v$ can be written as a direct sum of orthogonal subspaces,
$\mathcal{W}^{(v)}=\bigoplus_\mu \mathcal{W}^{(v)}_\mu$, which are
invariant under the action of all local terms in the Hamiltonian.
Moreover, denoting the neighbors of $v$ by $u_1, \ldots, u_D$, every
subspace $\mathcal{W}^{(v)}_\mu$ is (up to a local isometry) a
tensor product of ``sub-particles''
$\mathcal{W}^{(v)}_\mu=\mathcal{W}^{(vu_1)}_\mu\otimes\cdots
\otimes\mathcal{W}^{(vu_D)}_\mu$ such that the local Hamiltonian
associated with the edge $(v,u)$ acts non-trivially only on the
sub-particle $\mathcal{W}^{(vu)}_\mu$. Applying this decomposition
(namely the relevant isometries) to all particles, and projecting
into a specific $\mathcal{W}^{(v)}_\mu$ for every particle $v$ we
obtain a completely decoupled system in which each of the local
Hamiltonians acts on two isolated sub-particles, as illustrated in
\Fig{fig:BV}(b). 

Using this construction, Bravyi-Vyalyi \cite{ref:BV} find an
eigenbasis for $H$ made of trivial eigenvectors, each one generated
by a depth-$2$ quantum circuit. For example, a circuit generating a
groundstate is defined as follows: acting on an initial product
state, it first generates a (possibly entangled) state for each of 
the pairs of interacting sub-particles (as in~\Fig{fig:BV}). It then
applies the inverse of the relevant isometry on every $v$, namely,
on all sub-particles of $v$. This isometry rotates the state into
the $W^{(v)}_\mu$ subspace in which the desired groundstate lies.
Hence groundstates of $2$-local Hamiltonians are trivial; we
conclude that $2$-local $\CLH$s cannot be used to prove the $\NLTS$
conjecture.

The strong constraints placed on $\CLH$s by the BV lemma have many
consequences, not only on the structure of $2$-local but also to a
certain extent for $k$-local Hamiltonians. For instance, Aharonov
and Eldar~\cite{ref:DoritLior1} extended the lemma to $3$-local
Hamiltonians acting on $2$ or $3$-dimensional particles, showing
that these Hamiltonians cannot be used to prove the $\NLTS$
conjecture either.  Hastings~\cite{ref:NLTS1} extended it to
$k$-local $\CLH$s whose interactions hypergraphs are
``$1$-localizable'' (which, roughly, means that they can be mapped
to graphs continuously in a way that the preimage of every point is
of bounded diameter), as well as to very general planar
$\CLH$s~\cite{ref:Hastings2D}. Arad~\cite{ref:nogo} considered
slightly non-commuting $2$-local systems as perturbations of
commuting systems, showing that these systems have trivial
low-energy states as well.

One may ask whether trivial states can always be found for $\CLH$s
when the underlying interaction (hyper-)graph is a good enough
expander, by analogy with the Brand\~ao-Harrow result~\cite{ref:BH}
(\Sec{sec:BH}).  Indeed, Aharonov and Eldar \cite{ref:DoritLior2}
derived such a result in the commuting case for any constant
locality $k$. The results do not follow from \Ref{ref:BH} since the
notion of expansion for hypergraphs is very different from the
common notion for graphs. Let us therefore explain this notion.  We
associate with a Hamiltonian a bipartite graph in a natural way:
constraints are on one side and particles on the other; particles
are connected to the constraints they participate in.  The result of
\Ref{ref:DoritLior2} shows that if this bipartite graph is a good
small-set-expander (namely, for any set of particles of size $k$,
the set of constraints they are connected to has cardinality at
least $(1-\delta)$ times its maximal possible value $kD$, where $D$
is the number of constraints any particle participates in) then
trivial states exist below an energy which is of the order of
$\delta$. Just like in \Ref{ref:BV}, the better the expanders
(namely, the smaller $\delta$ is) the less appropriate the graphs
are for $\NLTS$. 

\subsection{Freedman-Hastings' construction}\label{sec:freedmanhastings}

We conclude this section by describing a result which provides a new
approach for proving the $\NLTS$ conjecture.  In a recent work,
Freedman and Hastings~\cite{ref:NLTS2} considered extending the
toric code from a $2$-dimensional grid to a graph which satisfies a
non-standard (and somewhat weaker than the one described above)
notion of hypergraph expansion, which they term
\emph{non-1-hyperfinite}.  We will not explain this result in detail
as it requires further additional topological and algebraic 
background; we will however attempt to give its flavor.

Think of the following question. We have seen in
Section~\ref{sec:commutingbad} that the toric code's groundstates
are non-trivial, but as soon as we allow the energy to increase,
trivial states exist. What would happen if we restricted our
attention to states which violate only \emph{one} type of operators,
say, the plaquette operators, while still requiring that all the
star operators have energy $0$? We refer to any local Hamiltonian
having no low-energy trivial states inside the groundspace of, say,
the star operators, as a ``one-sided error'' $\NLTS$.  Freedman and
Hastings noted that, although the toric code itself is not
``one-sided error'' $\NLTS$, it can be cleverly extended to a code
defined on a much better expanding graph whose groundstates do
satisfy this property.

Let us first see why the toric code groundstates are not one-sided
$\NLTS$.  Consider partitioning the lattice of the toric code in
$\ell\times \ell$ squares, for some constant $\ell$, removing only
the plaquette type operators acting on the corners of those squares.
It can be easily checked that the remaining Hamiltonian is $2$-local
provided one aggregates together all qubits in each square to one
big (but still of constant dimension) particle. The Bravyi-Vyalyi
lemma of Section~\ref{sec:commutingbad} can then be applied to derive
trivial states whose sole non-zero energy contribution comes from
the plaquette operators that were thrown away: this means that 
there are trivial states inside the groundspace of the star operators; 
the terms we threw 
away are of one type only, and hence the error is one-sided. 

\begin{figure}
  \begin{center}
      \includegraphics[scale=1.5]{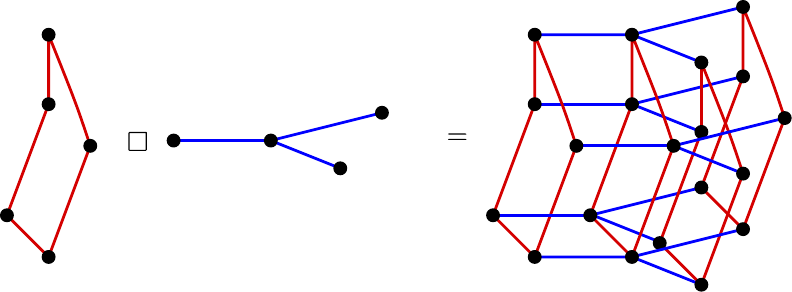}    
  \end{center}
  \caption{Cartesian product of graphs. We assume the girth of both
    graphs is strictly larger than $4$.  In this case all the
    plaquettes in the Cartesian product are defined by loops made of
    $4$ edges: two red edges (originating from the first graph) and
    two blue edges (originating from the second graph). Every such
    plaquette is associated with an $A$ term. \label{fig:cartesian}}
\end{figure}

To construct one-sided $\NLTS$ Hamiltonians, Freedman and Hastings
take a family of $D$-regular graphs $\{G_n\}$ over $n$ vertices,
with $D>2$ some fixed constant, and with diverging girth, and
consider their Cartesian product with themselves $\{G_n \square
G_n\}$.\footnote{The Cartesian product of $G_1=(V_1,E_1)$ with
$G_2=(V_2,E_2)$ is the graph whose vertices are $V_1\times V_2$ and
$(v_1,v_2)$ is connected to $(u_1,u_2)$ iff $(v_1,u_1)\in E_1$ and
$v_2=u_2$ or vice-versa -- see \Fig{fig:cartesian}.} If the girth of
the original graph is strictly larger than $4$, the result is a
graph in which each vertex is at the intersection of $4$-loops
(loops made of $4$ edges) that look as follows.  Start with a vertex
$(v_1,v_2)$, move along an edge $(v_1,u_1)$ in the first graph to
reach $(u_1,v_2)$; continue to move along an edge $(v_2,u_2)$ in the
second graph to reach $(u_1,u_2)$, then move again along $(v_1,u_1)$
to get to $(v_1,u_2)$ and finally again along $(v_2,u_2)$ to get
back to $(v_1,v_2)$ (see \Fig{fig:cartesian}). This leads to a
natural generalization of the toric code: we place qubits on the
edges, and identify the $4$-loops as ``plaquettes'', and the edges
adjacent to a vertex as ``stars''. We define $A$ terms as products
of $X$ operators over plaquettes, and $B$ terms as product of $Z$
operators over stars; this leads to the so-called homological code
on this graph \cite{ref:delgado}. 

Though the graphs $G_n$ are not necessarily expanders, the Cartesian
products $\{G_n\square G_n\}$ can be shown to possess a related
property: they are non-$1$-hyperfinite. Roughly, what this means is
the following.  We can consider $\{G_n \square G_n\}$ as a
$2$-simplicial complex. This is a two-dimensional object defined as
the union of the vertices of $\{G_n \square G_n\}$ (a zero-dimensional object),
 its edges (a one-dimensional object), and its
faces, corresponding to the plaquettes (a two-dimensional object).
Being non-$1$-hyperfinite means that it is impossible to
continuously map this two-dimensional object into a one-dimensional
object, namely, a graph, such that the pre-image of every point is
of a constant diameter; moreover, this is still impossible if one is
allowed to remove a constant fraction of the vertices of $\{G_n
\square G_n\}$ together with the plaquettes they participate in.
Freedman and Hastings then showed that for $\{G_n \square G_n\}$
there are no non-trivial state below a certain constant average
energy, as long as one is allowed to violate only one type of
operators.\footnote{We remark that Freedman and Hastings define
trivial states in a more restricted way than we did in
Subsection~\ref{sec:nontrivial}: they require that the
constant-depth circuit is \emph{local} with respect to some metric
that is derived from the interaction graph of the system. } 

It is interesting to see what the result of \Ref{ref:DoritLior2}
described in the previous subsection implies regarding the
construction of Freedman and Hastings~\cite{ref:NLTS2}, since the
bipartite graph corresponding to their construction can be
calculated and the small-set expansion error $\delta$ can be found: 
it turns out to be inverse polynomial in $D$, the degree of the
graph $G_n$. The result of \Ref{ref:DoritLior2} thus implies that
trivial groundstates of the Hamiltonians constructed
in~\cite{ref:NLTS2} exist below some constant average energy 
$\orderof{\delta}$, and this constant tends to $0$ as the small-set
expansion of the graphs underlying the construction \emph{improves}.
In similar spirit to the conclusion of Section~\ref{sec:BH}, one
cannot hope to improve the result of \cite{ref:NLTS2} and achieve
$\NLTS$ Hamiltonians for larger gaps by taking better and better
small-set expanders.

\section{Interactive proofs and qPCP}
\label{sec:qma-qproof}

In this last section of the survey we take a step back to take a
deeper look at the origins of the classical $\PCP$ theorem, rooted
at the notion of interactive
proofs~\cite{ref:Babai-IP,ref:Goldwasser-IP}, and ask if these
origins can inform our search for a quantum equivalent to the $\PCP$
theorem. Soon after their introduction, interactive proofs were
discovered to capture increasingly complex problems, from the
successive inclusions of coNP~\cite{GoldreichMW91},
PH~\cite{LunForKarNis92JACM} and finally $\PSPACE$~\cite{ref:Sha92}
in $\IP$, and culminating in the discovery of the surprising power
of multiple provers revealed by the equality
$\MIP=\NEXP$~\cite{BabForLun91CC}. It is the ``scaling down'' of the
latter result from $\NEXP$ to $\NP$, obtained by placing stringent
resource bounds on the verifier, that eventually led to the discovery of the $\PCP$
theorem.\footnote{We refer to~\Ref{ref:PCP-history} for a detailed
account of the fascinating history of the $\PCP$ theorem.} 

Just as it was instrumental in bringing forth the very notion of
locally checkable proofs, the perspective given by interactive proof
systems might also help shed light on the $\qPCP$ conjecture in its
proof verification form (Conjecture \ref{con:randomaccessqpcp}).  In
fact, attempting to directly approach the random access version of
the $\qPCP$, Conjecture~\ref{con:randomaccessqpcp}, may be easier
than tackling its gap amplification version,
Conjecture~\ref{con:qgap}: while the former follows from the latter
by a classical reduction, the other direction seems to require a
quantum reduction.

In this last section of the survey we turn to the study of quantum
interactive proofs with one or more provers. 
We start by describing an {\it exponential size} classical $\PCP$
for $\QMA$. The derivation of such exponential size $\PCP$s was an
important milestone on the route towards the $\PCP$ theorem
(\Ref{ref:AroraBarak}, chapter~$11$), and despite being much
simpler, they provide important intuition towards the polynomial
size proof. Moving on to the study of multiprover interactive
proofs, we ask whether the instrumental role played by these proof
systems in the proof of the $\PCP$ theorem can hint to a deeper
relationship between $\qPCP$ and interactive proofs with multiple
\emph{entangled} provers. Although we find that such a connection
does not seem to hold in the quantum case, the endeavor reveals
another interesting aspect of entanglement: the possibility of using
its nonlocal correlations to \emph{defeat} classical multiprover
interactive proof systems. This suggests further open questions
related to entanglement that we describe. 

\subsection{An exponential classical PCP for \boldmath{$\QMA$}}
\label{sec:qma-ip}

At first sight it might seem that devising an exponential-size
classical witness for an instance of LH is a trivial task: why not
simply use the classical description of the quantum witness, the
groundstate of the local Hamiltonian, as a list of $2^n$
coefficients (up to some precision), one per computational basis
state? The problem lies in efficiently {\it verifying} that this
state has low energy with respect to the Hamiltonian. Suppose for
example that one of the local terms is a projection of the first
qubit on the state $\ket{1}$.  The associated energy is the
probability to measure the first qubit of the witness in state
$\ket{1}$, the sum over all coefficients (squared) of computational
states labeled by a $'1'$ in the first position. Evaluating this sum
requires access to an exponential number of coefficients. As we can
see, depending on the choice of representation the locality of the
Hamiltonian may not correspond to locality in terms of the classical
witness. 

This difficulty suggests a different witness, for which the problem
mentioned above does not arise: the witness would be the list of all
local density matrices associated to the groundstate, on any subset
of $k$ qubits. Such a witness has \emph{polynomial} size, which
should make us suspicious. Indeed, the difficulty here is that
checking \emph{consistency} between different reduced local density
matrices, namely, that they come from a single quantum state, is
itself a $\QMA$-hard problem~\cite{ref:YiKai}! This approach seems
to be a dead end.\footnote{We note that consistency of density
matrices was shown to be $\QMA$-hard only under Turing reductions.
This is a weaker notion than the standard $\QMA$-hardness under Karp
reductions. Possibly, the above suggested witness {\it can} be used
in some way which does not require the full power of $\QMA$, but we
do not know how to do this.} 

However, the following line of argument shows that the ingredients 
required to devise such exponentially long proofs for $\QMA$ were
long known to exist. First, it is known that $\QMA \subseteq
\PSPACE$~\cite{KitaevW00,Vyalyi03}.  Second, $\PSPACE=\IP$
\cite{ref:Sha92} can be used to derive $\QMA\subseteq \IP$ and thus
any language in $\QMA$ has an efficient interactive proof. Next,
any interactive proof can be made into a static, exponentially
longer (but still efficiently verifiable) proof as follows.  The
static ``proof'' lists the whole tree of answers the prover would
give to the verifier's queries. A verifier checking the proof, based
on the first query that is made, partitions the proof into
sub-proofs, only one of which will be explored by subsequent
queries.  Finally, scaled-up versions of the $\PCP$
theorem~\cite{ref:PCP1,ref:PCP2} show that any such proof can be
encoded in a way that checking it only requires reading a constant
number of bits. Although this requires the use of heavy-handed
tools, such as low-degree tests, the parallel repetition theorem,
and composition of verifiers, it does prove that exponentially long
classical $\PCP$s for $\QMA$ exist.  We feel, however, that it is
insightful to ``open'' the above line of argument and see what it
may teach us in relation to $\qPCP$. In the following subsections we
describe an interactive proof system for $\QMA$, from which an
exponentially long proof can be derived as above. The resulting
proof is efficiently verifiable (it too can be turned into a proof
whose checking requires reading only a constant number of bits,
using the same classical methods mentioned above.) It is worthwhile
noting that the resulting proof does not explicitly encode the
quantum witness, but instead directly works with the local
Hamiltonian. In a sense, our proof is a step away from natural
attempts at quantizing the classical $\PCP$ transformations; taking
such a step is motivated in part by Corollary~\ref{coro:BH}, which
indicates that the most ``natural'' transformations may fail in the
quantum setting. 

\subsubsection{Reducing to a trace computation}
\label{sec:traces}

For simplicity, let us assume that we are trying to distinguish
between the groundstate energy of a given local Hamiltonian
$H=\sum_{i=1}^m H_i$ being $0$ or at least $\Gap = 1/\poly(n)$. Our
first step consists of amplifying this gap algebraically. Define
$M\EqDef \Id-H/m$, and consider the operator $M^\ell$ for an integer
$\ell = \Omega(mn/\Gap)$.\footnote{This definition essentially
corresponds to repeating the $\QMA$ verifier's procedure for LH (see
\Sec{sec:cooklevin}) \emph{sequentially} a polynomial number of
times applied on the \emph{same} register.  Here we are taking
advantage of our assumption on the groundstate energy being $0$;
Marriott and Watrous~\cite{MarriottW05qma} showed that a similar
``sequential'' amplification, repeatedly re-using the same $\QMA$
witness, can also be performed in general.} If $H$ has ground energy
$0$ then the largest eigenvalue of $M^\ell$ is $1$, whereas if it is
at least $\Gap$ then the largest eigenvalue of $M^\ell$ is at most
$2^{-n-1}$. In order to distinguish between the two, it will suffice
to compute the trace of $M^\ell$: in the first case it is at least
$1$, while in the second it is at most $2^n\cdot 2^{-n-1} = 1/2$.

Since $M^\ell$ is an exponential size matrix, computing its trace
requires evaluating an exponential-size sum.  We note that computing
the trace of the $k$-local Hamiltonian $H$ (and therefore of $M$),
despite it being an exponential size matrix, can be easily done as
$\Tr H = \sum_{i=1}^m \Tr H_i$, and $\Tr H_i$ is easy to calculate:
it equals $2^{n-k}$ times the trace of the local term, a matrix of
constant size $2^k\times 2^k$.  The reason it is difficult to
compute the trace of $M^\ell$ is that it involves high powers of
$H$, which eliminate its local nature. Hence what we need to show is
that, though hard to compute directly, a claimed value for
$\Tr(M^\ell)$ can be efficiently \emph{verified} through an
interactive proof system involving only local computation from the
verifier.

\subsubsection{A first interactive proof for the trace}

We first handle a simplified version of our problem: we consider a 
Hermitian matrix $A$ (which we will eventually take to be $M^\ell$),
such that it is possible to efficiently compute any diagonal entry
of $A$ in the standard basis, i.e., expressions of the form
$A_{i_1\ldots i_n}\EqDef\bra{i_1\ldots i_n}A\ket{i_1\ldots i_n}$,
where $i_1,\ldots, i_n \in \{0,1\}$ and $\ket{i_1 \ldots i_n}\EqDef
\ket{i_1}\otimes\cdots\otimes\ket{i_n}$.  We note that for
$A=M^{\ell}$, computing a single diagonal entry requires the
multiplication of exponential size matrices, which is a priori
computationally hard; we shall return to this issue later. Using
this notation, our goal is now to verify the exponential sum $\Tr(A)
= \sum_{i_1,\ldots, i_n} A_{i_1\ldots i_n}$.

Taking inspiration from the proof that $\PSPACE\subseteq\IP$, but
simplifying to the maximum, a first na\"ive attempt at giving an
interactive proof for the statement ``$\Tr(A)=a$'' could be as
follows. The verifier will ask the prover to provide answers to
random questions; based on the prover's answers, he will ask further
questions.  Since the prover does not know the next question when
answering the current one, he needs to answer in a way consistent
with all (or most) possible future questions. The protocol is
designed so that if he is cheating, it will be all but impossible
for him to succeed in doing so. Let $A_{i_1}\EqDef \sum_{i_2,\ldots,
i_n} A_{i_1\ldots i_n}$ denote the partial sum, so that $\Tr(A) =
\sum_{i_1} A_{i_1} = A_0 + A_1$.  The verifier first asks the prover
for $A_0$ and $A_1$ and verifies that $A_0 + A_1 = a$. He then flips
a fair coin $c_1\in\{0,1\}$, and asks the prover for $A_{c_1,0}$ and
$A_{c_1,1}$, where $A_{c_1,i_2}\EqDef \sum_{i_3,\ldots, i_n} A_{c_1,
i_2, i_3\ldots i_n}$. He verifies that $A_{c_1} = A_{c_1,0} +
A_{c_1,1}$, and continues this way recursively for $n$ steps. In the
end, he arrives at an expression of the form $A_{c_1,\ldots,c_n}$,
which he can evaluate by himself.

This procedure, however, is not sound: the prover can
cheat and be caught with only exponentially small probability. For
example, assume all diagonal entries are $1$ except for one of
them, which is $2$. The correct trace is $2^{n}+1$, but the prover
wants to convince the verifier the trace is $2^n$.  To achieve this, the
prover could declare $a=2^n$.  To be consistent, in the next step he
declares $A_0=A_1=2^{n-1}$.  One of those values, say $A_0$ (this
depends on the location of the single `$2$' entry in the matrix)
corresponds to the correct trace of the corresponding block matrix,
here $A_0$; but the other does not.  If the verifier chose the value
$c_1=0$, the prover is safe and can be truthful for the remainder of
the protocol. Otherwise, he is back to the previous scenario.  Thus, at
each iteration, the prover has probability half to ``escape'' and
can proceed truthfully from there; he is only caught if no round gave him
the possibility of escaping, which happens with probability $2^{-n}$. 

The main ingredient in the $\IP=\PSPACE$ proof, namely, the
sum-check protocol~\cite{LunForKarNis92JACM}, suggests a particular
way around this issue, reducing the probability of the cheating
prover to escape from very close to $1$ to exponentially small.
Roughly, the idea is to introduce \emph{redundancy}.
In our 
particular context, this would be achieved by
defining a multilinear polynomial whose variables are the $n$
Boolean bits specifying a coordinate in the matrix, and whose
values are the corresponding diagonal elements of $A$. 
Redundancy is introduced by extending the field over which the
polynomial is defined from $\mathbb{F}_2$ (the values of the coin in
the protocol above) to the finite field $\mathbb{F}_p=
\{0,\ldots,p-1\}$, where $p\gg 2^{n}$.  The verifier is assumed to
be able to evaluate the polynomial at a random point $x\in
\mathbb{F}_p^n$.  The idea is that soundness follows from the fact
that two such polynomials (of total degree at most $n$) cannot agree
on too many values without being equal.

\subsection{A sum-check protocol for quantum systems}

Here we introduce a somewhat different form of redundancy. While
conceptually it plays the same role, we feel that it is more natural
in the present setting, and in particular respects the intrinsic
``locality'' of the problem.  Instead of adding redundancy through
the consideration of linear combinations of the diagonal entries of
$A$ (as would result from the extension to $\mathbb{F}_p$ which is
done in the classical sum-check protocol sketched above), our proof
system enables the verifier to compute the sum of all diagonal
entries of $A$ in \emph{any} (suitably discretized) \emph{product}
basis, given the ability to directly evaluate any \emph{one} such
entry.  Assume henceforth that the verifier has the ability to
efficiently compute the value of $
A_{\psi_1\cdots\psi_n}\EqDef\bra{\psi_1\cdots\psi_n}
A\ket{\psi_1\cdots\psi_n}$, where $\ket{\psi_1\cdots \psi_n}$ is
shorthand for $\ket{\psi_1}\otimes \cdots \otimes \ket{\psi_n}$ and
$\ket{\psi_i}$ are single qubit states taken from some suitable
discretization of the Hilbert space of a single qubit. 

We first introduce some notation. For
a set of single qubit states $\ket{\psi_1}, \ldots, \ket{\psi_r}$, let
\begin{align*}
  \Pi_{\psi_1,\ldots, \psi_r} \EqDef \ket{\psi_1}\bra{\psi_1}\otimes
    \cdots \otimes\ket{\psi_r}\bra{\psi_1} \otimes \Id_{n-r}
\end{align*}
denote the projection into the subspace in which the first $r$
qubits are in the pure state
$\ket{\psi_1}\otimes\cdots\otimes\ket{\psi_r}$, and define
\begin{align*}
  A_{\psi_1\ldots\psi_r}\EqDef \Tr(A\Pi_{\psi_1,\ldots, \psi_r}) \ .
\end{align*}
When $\ket{\psi_j}$ are taken
from the computational basis, i.e.,
$\ket{\psi_j}\in\{\ket{0},\ket{1}\}$, we recover the previous
definition of $A_{i_1,\ldots,i_r}$. Next, we define the partial
trace $\Tr_{\setminus i}$ to be the tracing out of all qubits except
for the $i$th one (see \Sec{sec:multi} for the definition of a partial
trace). Note that for any matrix $B$, $\Tr_{\setminus i}(B)$ is a
$2\times 2$ matrix, and that $\Tr(B) = \Tr_i \Tr_{\setminus
i}(B)=\Tr_{\setminus i} \Tr_i(B)$.

Our interactive proof for ``$\Tr(A)=a$'' proceeds as follows. At the
first step, the verifier asks the prover for the $2\times 2$
matrix $a^{(1)}\EqDef \Tr_{\setminus 1} (A)$ (this corresponds to
the reduced matrix on the first qubit). He verifies that $\Tr
a^{(1)} = a$. Then, in the second step, he selects a random state
$\ket{\psi_1}$ and asks the prover for the $2\times 2$ matrix
$a^{(2)}\EqDef \Tr_{\setminus 2} (A\Pi_{\psi_1})$ (this corresponds
to projecting the first qubit on the state $\ket{\psi_1}$, and
asking for the reduced matrix on the second qubit.).  He then
verifies the statement that $\Tr(a^{(1)}\ket{\psi_1}\bra{\psi_1}) =
A_{\psi_1}$ by checking that $\Tr(a^{(1)}\ket{\psi_1}\bra{\psi_1}) =
\Tr a^{(2)}$. At the $i$th step, he chooses a random state
$\ket{\psi_{i-1}}$, asks the verifier for $a^{(i)}\EqDef
\Tr_{\setminus i} (\Pi_{\psi_1\ldots\psi_{i-1}} A)$, and verifies
the statement ``$\Tr(a^{(i-1)}\ket{\psi_{i-1}}\bra{\psi_{i-1}}) =
A_{\psi_1\ldots \psi_{i-1}}$'', by checking that
\begin{equation}
\Tr(a^{(i-1)}\ket{\psi_{i-1}}\bra{\psi_{i-1}}) = \Tr a^{(i)}. 
\end{equation}
Finally after the $n$th step, he chooses a random state
$\ket{\psi_n}$ and verifies the statement
``$\Tr(a^{(n)}\ket{\psi_n}\bra{\psi_n}) = A_{\psi_1\ldots \psi_n}$''
by directly calculating the right-hand side by himself.

\subsection{Soundness analysis}

Why is this protocol sound?  Let us assume the prover is cheating
and consider some run of the protocol.  The verifier asks the prover
$n$ questions, and receives $n$ answers in the form of $2\times 2$
matrices $a^{(i)}$, which are supposed to be equal to
$\Tr_{\setminus i}(\Pi_{\psi_1\ldots\psi_{i-1}}A)$. To fool the
verifier, the prover must at some point start giving the verifier
the true matrices, or somehow pass the last test with a wrong matrix
for $a^{(n)}$. We will show that his chances of succeeding in either
case are slim. Assume first that at the $i$th step he gives the
verifier $a^{(i)} \neq \Tr_{\setminus
i}(\Pi_{\psi_1\ldots\psi_{i-1}}A)$, and in the next step he gives
the true matrix, {$a^{(i+1)}=\Tr_{\setminus
(i+1)}(\Pi_{\psi_1\ldots\psi_{i}}A)$}.  The verifier then compares
$\Tr(a^{(i+1)}) =\Tr(\Pi_{\psi_1\ldots\psi_i}A)$ to
$\Tr(\ket{\psi_i}\bra{\psi_i}a^{(i)})$. The difference between these
two numbers can be written as
\begin{align*}
  \Tr(\Pi_{\psi_1\ldots\psi_i}A) -
  \Tr(\ket{\psi_i}\bra{\psi_i}a^{(i)})
  &= \Tr_i\big[ \ket{\psi_i}\bra{\psi_i}
    (\Tr_{\setminus i} \Pi_{\psi_1\ldots\psi_{i-1}}A -
    a^{(i)})\big] \\
  &= \Tr_i( \ket{\psi_i}\bra{\psi_i} \Delta ) 
    = \bra{\psi_i}\Delta\ket{\psi_i}\ ,
\end{align*}
where $\Delta\EqDef \Tr_{\setminus i} \Pi_{\psi_1\ldots\psi_{i-1}}A
- a^{(i)}$ is a $2\times 2$ matrix. So the prover will only be able
to convince the verifier with an honest answer if
$\bra{\psi_i}\Delta\ket{\psi_i}=0$. Since by assumption, $\Delta\neq
0$ and $\ket{\psi_i}$ is a random state, then clearly, if we were
working with exact arithmetic, the chances for that to happen would
have been zero. Similarly, suppose the prover gives the verifier
{$a^n\neq \Tr_{\setminus n}\Pi_{\psi_1\ldots\psi_{n-1}}$}.  The
probability that he passes the verifier's test 
``$\Tr(a^{(n)}\ket{\psi_n}\bra{\psi_n}) = A_{\psi_1\ldots \psi_n}$''
is zero (with exact arithmetic) since for a random $\ket{\psi_n}$,
with probability $1$ it will hold that 
{$\Tr(a^{(n)}\ket{\psi_n}\bra{\psi_n}) \neq \Tr(\Tr_{\setminus
n}\Pi_{\psi_1\ldots\psi_{n-1}}\ket{\psi_n}\bra{\psi_n})= 
A_{\psi_1\ldots\psi_n}$}. 

We note that additional care must be taken to conclude the proof
since all numbers must be represented with finite precision.  In
this case, the probabilities mentioned above for the cheating 
prover to pass the test do not vanish; however one can verify 
(using straightforward though somewhat subtle arguments that we
omit here) that each of them can be made exponentially small by
performing all computations with a fixed polynomial number of bits
of precision.

\subsubsection{The final interactive proof}

We are almost done with the proof, but for one difficulty --- we do
not know how to compute diagonal coefficients of $M^\ell$, even in a product
basis, efficiently. Nevertheless, we do know how to evaluate
expressions such as
$\bra{\psi_1\ldots\psi_n}M\ket{\psi_1\ldots\psi_n}$, because
$M=(\Id-\frac{1}{m}\sum_i H_i)$, and we can efficiently calculate
$\bra{\psi_1\ldots\psi_n}H_i\ket{\psi_1\ldots\psi_n}$ for $k$-local
terms such as $H_i$. We must therefore find a way to ``break'' $M^\ell$
into single powers of $M$. Luckily, this task is not very different from
computing the trace, and we can perform it by using the previous
protocol iteratively. To verify the statement
\begin{align}
\label{eq:statement}
  \bra{\psi_1\ldots\psi_n} M^\ell 
    \ket{\psi_1\ldots\psi_n} = A_{\psi_1\ldots\psi_n} ,
\end{align} 
for some given real number $A_{\psi_1\ldots\psi_n}$, we write
$\bra{\psi_1\ldots\psi_n}M^\ell \ket{\psi_1\ldots\psi_n} = \Tr
(\Pi_{\psi_1\ldots\psi_n} M^\ell)$, which using the cyclic property
of the trace equals $\Tr (M\Pi_{\psi_1\ldots\psi_n} M^{\ell-1})$. In
this form~\eqref{eq:statement} can be verified by invoking the
previous protocol again, resulting in the requirement to verify the
value of
\begin{align*}
  \Tr (\Pi_{\tilde{\psi}_1\ldots\tilde{\psi}_n}
    M\Pi_{\psi_1\ldots\psi_n} M^{\ell-1})  \ ,
\end{align*}
where $\ket{\tilde{\psi}_1}, \ldots, \ket{\tilde{\psi}_n}$ are the
random states chosen in the second application of the protocol.
Proceeding this way $\ell$ times, we end up having to verify the
value of an expression of the form $\Tr (\Pi_\ell M \Pi_{\ell-1} M
\cdots \Pi_1 M)$, where $\Pi_j\EqDef
\Pi_{\psi_1^{(j)}\ldots\psi_n^{(j)}}$ is the projection used at the
end of the $j$-th application of the protocol, and
$\ket{\psi_1^{(j)}},\ldots,\ket{\psi_n^{(j)}}$ are the random states
that define it. It is now easy to see that $\Tr (\Pi_\ell M
\Pi_{\ell-1} M \cdots \Pi_1 M)$ can be written as the product 
$M_{\ell, \ell-1}\cdot M_{\ell-1, \ell-2} \cdots M_{2,1} \cdot
M_{1,\ell}$, where 
\begin{align*}
   M_{i,j} \EqDef \bra{\psi^{(i)}_1\ldots \psi^{(i)}_n} M 
   \ket{\psi^{(j)}_1\ldots \psi^{(j)}_n} \ .
\end{align*}
This is a local expression that can be evaluated efficiently,
finishing the proof.

\subsection{The two-provers angle}\label{sec:two-provers}

We end this section by focusing on an idea which played a very
important role in the path towards the $\PCP$ theorem: its direct
correspondence with \emph{multiprover} interactive proofs. To see
the connection, let us take the example of $k$-$\SAT$ (though any
$k$-$\CSP$ would do). The $\PCP$ theorem implies that any $k$-$\SAT$
formula $\varphi$ can be transformed into another formula $\varphi'$
over polynomially many more variables such that, if $\varphi$ is
satisfiable then so is $\varphi'$, but if $\varphi$ is not
satisfiable then at most $99\%$ of the clauses of $\varphi'$ can be
simultaneously satisfied. Hence there is an efficiently verifiable
proof for the satisfiability of $\varphi$: simply ask for an
assignment to the variables of $\varphi'$ satisfying as many clauses
as possible. The verifier will pick $200$ clauses at random, read
off the corresponding at most $200k$ variables from the proof, and
accept if and only if the assignment satisfies all $200$ clauses.

We can make this proof checking procedure into a two-prover
interactive proof system as follows: the verifier chooses a clause
of $\varphi'$ at random, and asks a first prover for an assignment
to its variables. He also chooses a single one of the variables
appearing in the clause, and asks the second prover about it. He
accepts if and only if the first prover's answers satisfy the
clause, and the second prover's answer is consistent with that of
the first. It is this consistency check that allows to relate any
strategy of the provers to a proof: since the second prover only
ever gets asked about single variables, his strategy \emph{is} an
assignment to the variables.\footnote{ Although the provers may a
priori use randomized strategies, including the use of shared
randomness, it is not hard to see that this cannot help: in the
classical setting we may always restrict attention to deterministic
strategies.} Consistency with the first prover, together with
satisfaction of the first provers' answers, implies that, provided
the provers are accepted with high probability, the second provers'
assignment must satisfy most clauses of $\varphi'$.

Interestingly, this correspondence between locally checkable proofs
and multiprover interactive proof systems completely breaks down in
the case of quantum proofs. To see why, let's try to make an
instance of the $k$-local Hamiltonian problem into a two-prover
interactive proof system in a similar manner as we did for instances
of $k$-SAT. The natural idea would be to ask each prover to hold a
copy of the groundstate $\ket{\Psi}$. The referee would then choose
one of the terms $H_i$ of the Hamiltonian at random, and ask the
first prover to hand him the $k$ qubits on which $H_i$ acts. He
would then choose one of these $k$ qubits at random, and ask the
second prover to provide it.  Upon receiving the qubits, the
verifier can either evaluate the energy of these qubits with respect
to $H_i$ or check consistency between the second prover's qubit and
the matching qubit from the first prover's answer...or can he? There
is a serious issue with this procedure of course, an issue we
encountered many times before. Because of the possible presence of
entanglement in the groundstate, it could be that the states held by
cheating provers are very different (even just on $k$ of the qubits)
but, on any one qubit, they are identical: it is impossible to check
consistency between each provers' ``proofs'' by a local
procedure.\footnote{There are many other difficulties: for instance,
we have no guarantee that, when asked for qubits $i$ or $j$, the
second prover actually sends us distinct qubits that can be
``patched'' into a single global state. An even more basic problem
is that there is no quantum procedure that can decide equality of
arbitrary quantum states with good success probability (even by
acting globally) --- such a procedure only exists for the case of
pure, not mixed, states.}

We are stuck -- there does not seem to be a straightforward notion
of a quantum multiprover interactive proof that would capture the
$\qPCP$, as formulated in Conjecture~\ref{con:randomaccessqpcp}, as
classical interactive proof systems capture the classical $\PCP$.
Nevertheless, our attempts provided a glimpse of a new model of
interactive proofs in which the provers may be entangled.  Formally,
the corresponding class, $\MIP^*$, was defined by Cleve et
al.~\cite{CHTW04} as the class of languages that can be verified by
a classical polynomial-time verifier with the help of two
all-powerful, untrusted but non-communicating, quantum entangled
provers.\footnote{One could also consider the class $\QMIP^*$ in
which the verifier is also allowed to be quantum, and exchange
quantum messages with the provers. However, it was recently shown
that $\QMIP^*=\MIP^*$~\cite{ReichardtUV13leash}. Just as for
single-prover interactive proofs, in which
$\QIP=\IP$\cite{JainJUW10qip}, the use of quantum messages does not
bring additional power in this setting.}
 
What can be said about this class?  Interestingly, the classical
inequality $\MIP=\NEXP$ does not carry any non-trivial implication
for the power of $\MIP^*$, in neither direction.  First, recall that
the easy direction, $\MIP\subseteq \NEXP$, is obtained by arguing
that optimal deterministic strategies, pairs of functions from
questions to answers, can be guessed in non-deterministic
exponential time. However, in the quantum case there is no a priori
bound on the dimension, or complexity, of optimal (or even
approximately optimal) quantum strategies for the provers. In fact,
no upper bound is known for $\MIP^*$: proving such a bound is a
major open question in entanglement theory.  Turning to the reverse
inclusion $\NEXP \subseteq \MIP$, the reasons it does not carry over
in any automatic way go back to foundational work of John Bell in
the $1960$s~\cite{Bell:64a} related to the EPR states mentioned in
Section~\ref{sec:multi}. Bell's main observation (following
Einstein, Podolsky and Rosen~\cite{epr}) is that, while the quantum
provers do not communicate, they can still use their entangled state
in a meaningful way to generate correlations (\emph{joint}
distributions on their answers to the verifier's queries) that
\emph{cannot} be simulated using shared randomness alone. These
additional correlations may (and, in some cases, such as unique
games~\cite{KRT08unique}, do) break the soundness of existing proof
systems. In a recent result~\cite{ItoV12mipstar} Ito and Vidick 
showed that Babai et al.'s multilinearity test, a key test in their
multiprover interactive proof system for $\NEXP$, is \emph{robust}
to entanglement: the test remains sound even when executed with
quantum provers. Using this they showed that $\NEXP\subseteq
\MIP^*$: allowing the provers to be entangled does not \emph{reduce}
the expressivity of the class $\MIP^*$. 

Despite the difficulty in directly connecting the multiprover model 
to the quantum $\PCP$ conjecture as formulated in
Conjecture~\ref{con:randomaccessqpcp}, one may still be able to
derive interesting results regarding $\qPCP$ from the study of
$\MIP^*$.  Such a result was recently derived by one of
us~\cite{ref:Vid13}, showing a \emph{different} quantum equivalent
to the classical $\PCP$ theorem, using its formulation in terms of
multiplayer games. Classically, this formulation states that it is
$\NP$-hard to determine whether some $2$-player games (such as the
$k$-SAT game described above) can be won with probability $1$, or at
most some constant strictly smaller than $1$. Ref.~\cite{ref:Vid13}
proves that $\NP$-hardness still holds even if the players are
quantum and are allowed to be entangled. As discussed above, even
$\NP$-hardness (instead of, say, $\QMA$-hardness) is not trivial
here: since quantum provers may achieve higher success
probabilities, existing protocols for $\NP$-hard languages may no
longer be sound.  The proof builds on showing that the low-degree
tests, which extend the multilinearity test and were used in the
original proof of the $\PCP$ theorem~\cite{ref:PCP1,ref:PCP2},
possess strong ``robustness'' properties -- namely they are sound
even when the provers are allowed to use entanglement. 

\section{Concluding remarks} 

This column contains a wealth of open problems, which we will not
repeat here. We highlight several problems not encountered in the
text: 

\begin{itemize}
  \item $\QMA$. Perhaps there is a better characterization of $\QMA$  
    that would make the $\qPCP$ conjecture more accessible? That the
    current formulation is not optimal might be hinted to by the
    fact that many natural questions on this class are still open.
    These include the analogue of the Valiant-Vazirani theorem
    \cite{ref:ValiantVazirani}, which was only partially answered in
    \Ref{ref:uniqueQMA1, ref:uniqueQMA2}, the question of whether
    one sided and two sided error are equivalent
    \cite{ref:AaronsonQMA1,ref:onesidedQMA, ref:onesidedQCMA}, the
    question of whether classical and quantum witnesses are
    equivalent (e.g., \cite{ref:qcma.kuperberg}), the power of the
    class when the witness is promised to be made of two or more
    unentangled registers instead of one register (see, e.g.,
    \cite{ref:unentanglement,ref:unentanglement2}), and more.  
    
  \item Quantum locally testable codes ($\qLTC$s). 
    A locally testable code ($\LTC$) guarantees that if a word is
    very far from the code, a random local test will detect it with
    good probability (this is referred to as the code being
    ``robust'').  $\LTC$ codes such as the Hadamard and the long
    code have played a crucial role in the classical $\PCP$ proof
    \cite{ref:PCP1, ref:PCP2}; in fact, one can show that any 
    $\PCP$ {\it of proximity}, which is a weak version of $\PCP$
    \cite{ref:PCPP} implies an $\LTC$ with related parameters. The
    quantum version of $\LTC$s was defined in \Ref{ref:DoritLior3}
    where weak bounds on the robustness of any $\qLTC$ were given.
    Many questions arise, e.g., what is the optimal robustness of
    $\qLTC$s? could it be constant?  How do the parameters of
    classical and quantum $\LTC$s relate?  The notion of $\qLTC$
    seems to be related \cite{ref:DoritLior3} to the $\NLTS$
    conjecture,  though no
    formal connection is known.  As was explained at the end of
    Section~\ref{sec:quantize}, quantum error correcting codes cannot be
    used in the most straightforward manner to achieve $\qPCP$s; the
    connection between this notion and $\qPCP$ is yet to be
    clarified.   

  \item Efficient description of quantum states. Even if the
    $\NLTS$ conjecture holds, one would need a stronger version of
    it to be relevant for the $\qPCP$: in fact, the $\qPCP$ requires
    that the low energy states of the output Hamiltonians cannot
    have a polynomial classical description from which their energy
    can be efficiently computed classically. Non-triviality of the
    states is not sufficient to ensure this; for example, Aguado and
    Vidal~\cite{ref:AguadoVidal} use a graphical construction 
    called MERA to provide an efficient classical description for
    the toric code states, which, as we have seen, are non-trivial. 
    MERA are an instance of a major thread in quantum Hamiltonian
    complexity: that of finding graphical descriptions called {\it
    tensor networks} for physically interesting classes of quantum
    states, such as groundstates of gapped Hamiltonians.
    Understanding the conditions under which such efficient tensor
    networks exist is thus of high relevance; the groundstates of
    the Hamiltonians used in any $\qPCP$ construction must not have
    such descriptions. 

    We mention a relevant work of Schuch~\cite{ref:schuch}. Schuch
    describes a classical witness for the fact that the ground energy
    of a commuting Hamiltonian on a square lattice of qubits, such
    as the toric code, is zero. However, this is done not by
    providing an explicit classical description of the groundstate,
    but rather in a much more indirect way. Very roughly, Schuch writes
    the Hamiltonian as the sum of two $2$-local Hamiltonians; he can
    then use the Bravyi-Vyalyi machinery~\cite{ref:BV} with respect
    to each one of those, to construct a certificate which shows
    that the \emph{intersection} of their two groundspaces is
    non-zero.  Of course, if the $\qPCP$ conjecture is true, then the
    Hamiltonians in the construction cannot allow such indirect
    witnesses either.

  \item Entangled provers. As we saw in Section~\ref{sec:two-provers}, 
    the tight connection that exists in the classical setting
    between locally checkable proofs, constraint satisfaction
    problems and multiprover interactive proofs breaks down in the
    quantum setting: locally checkable proofs can no longer be
    consistently shared between multiple provers. Although the formal
    connection seems to be lost, it may nevertheless be fruitful to
    examine questions on the one aspect in light of progress on the
    other. Does the recent proof \cite{ItoV12mipstar} that
    $\NEXP\subseteq \MIP^*$ have any implications for $\qPCP$? Could
    it be that $\QMA_{\textsc{EXP}}$, the exponentially scaled-up
    analogue of $\QMA$, is included in $\MIP^*$? Indeed, no
    \emph{upper} bounds on the latter class are known --- none at all.

\end{itemize}

\section*{Acknowledgments} 

The authors would like to thank Fernando Brand{\~a}o, Lior Eldar,
Aram Harrow and Matt Hastings for very useful comments on this
manuscript.


\bibliographystyle{alphaabbrvprelim}

{~}

\bibliography{qpcprefs}

\end{document}